\documentstyle[11pt,preprint,aps,epsfig]{revtex}
\tightenlines
\begin{document}
\draft
\title{Fundamental Framework for Technical Analysis}
\newcommand{\Dlt}{\Delta}
\newcommand{\lbd}{\lambda}
\newcommand{\sgm}{\sigma}
\newcommand{\vp}{\varphi}
\newcommand{\be}{\begin{equation}}
\newcommand{\ee}{\end{equation}}

\author{J. V. Andersen$^1$, S. Gluzman$^2$ and D. Sornette$^{1,3}$\\
$^1$ Nordic Institute for Theoretical Physics\\
Blegdamsvej 17, DK-2100 Copenhagen, Denmark\\
$^2$ Laboratoire de Physique de la Mati\`{e}re Condens\'{e}e\\ CNRS UMR6622
and Universit\'{e} de Nice-Sophia Antipolis\\ B.P. 71, Parc
Valrose, 06108 Nice Cedex 2, France \\
$^3$ Institute of Geophysics and
Planetary Physics\\ and Department of Earth and Space Science\\
University of California, Los Angeles, California 90095\\}
\maketitle

\begin{abstract}
Starting from the characterization of the past time evolution of market
prices in terms of two fundamental indicators, price velocity and price
acceleration, we construct a general classification of the possible patterns
characterizing the deviation or defects from the
random walk market state and its time-translational
invariant properties. The classification relies on two
dimensionless parameters, the Froude number characterizing the relative
strength of the acceleration with respect to the velocity and the time
horizon forecast dimensionalized to the training period. Trend-following and
contrarian patterns are found to coexist and depend on the
dimensionless time horizon. The classification is based on the symmetry
requirements of invariance with respect to change of price units and of
functional scale-invariance in the space of scenarii. This
``renormalized scenario'' approach is
fundamentally probabilistic in nature and exemplifies the view that multiple
competing scenarii have to be taken into account for the same past history.
Empirical tests are performed on
about nine to thirty years of daily returns of twelve data sets
comprising some major indices (Dow Jones, SP500, Nasdaq,
DAX, FTSE, Nikkei), some major bonds (JGB, TYX) and some major currencies
against the US dollar
(GBP, CHF, DEM, JPY). Our ``renormalized scenario'' exhibits
statistically significant predictive power in essentially all market phases.
In constrast, a trend following strategy and trend + acceleration
following strategy perform well only on different and specific market phases.
The value of
the ``renormalized scenario'' approach lies in the fact that it
always finds the best of the two, based on a calculation of the stability of
their predicted market trajectories.

\end{abstract}

\section{Introduction}

There is increasing evidences that even the most competitive markets are not
strictly efficient \cite{Farmer,Zhang}. In particular, a set of studies in
the academic finance literature have reported anomalous earnings which
support technical analysis strategies \cite{Joy,Jagadeesh,Lehmann,Brocketal}
(see \cite{Murphy} for a different view). A recent study of 60 technical
indicators on 878 stocks over a 12-year period \cite{Bauer} finds that the
trading signals from technical indicators do on average contain information
that may be of value in trading, even if they generally underperform
(without taking due consideration to risk-adjustments of the returns) a
buy-and-hold strategy in a rising market by being relatively rarely invested.

Another class of studies views the market place as a complex self-organizing
system \cite{Santefe,Farmer} which
suggests that technical analysis may have some value. This approach views
the traders develop strategies both fundamental, technical and any mixture
of them, which
adapt and react to the pattern these strategies create by their collective
action.
This concept is common
to other systems in Nature as well, including ions in a spin glass,
cells in an immune system, faults in the crust, etc.
In the economy, economic agents, banks,
consumers, firms, or investors, continually adjust their market moves, buying
decisions, prices, and forecasts to the situation these moves or decisions or
prices or forecasts together create. The challenge is to understand how these
actions, strategies, and expectations react to and
change with the aggregate patterns these create. Since the self-organization
is not instantaneous via the processes of adjustment and change as the
traders react,
the market evolves leading to a novel ajustment of the traders. The market as
a complex system is a process
that constantly evolve and unfold over time. It might thus
exhibit some degree of predictability in contrast to the efficient market
hypothesis
and the proof that correctly anticipated prices are random \cite{Samuelson}.

Most trading systems fall into two classes.

\begin{itemize}
\item  The first one is trend-following: the technical indicators attempt to
detect a significant trend and issue a signal for the trader to profit from
the trend.

\item  The second class is contrarian: the technical indicators try to
measure a change of trend. For instance, the oscillator indicators, which
are used to model the cyclical nature of markets, typically will filter the
trend out of prices, leaving only the remaining changes of trends.
\end{itemize}

Notwithstanding their multiple forms and sometimes complicated formulations
\cite{Achelis,techn}, technical indicators can be seen as reducing
essentially to combinations of measurements of

\begin{enumerate}
\item  a price velocity $v$, defined as the rate of change of the price
(possibly over different time scales), and of

\item  a price acceleration $g$, defined as the rate of change of the price
velocity.
\end{enumerate}

The velocity is a measure of the strength of a trend, while the acceleration
quantifies its persistence. For instance, when a market has been trending
upward and then begins to decelerate, an oscillator indicator will level
off, suggesting an approaching market top. Likewise, if a market is trending
downward and this trend decelerate, then a bottom would be forecasted by the
technical indicators. This phenomenology has suggested qualitative analogies
with Newton's law of classical motion, according to which velocities (or
momentum) change because market forces are exerting their influence and
produce acceleration/deceleration.

This view is in contrast to the (weak) efficient market and random walk
hypotheses, deeply ingrained in the financial academic literature, according
to which future variations of prices are unpredictable (at least from the
sole knowledge of past prices). Here, we develop a fundamental theory of
technical analysis based on the idea that trading signals, when they can be
identified, quantify local deviations from or equivalently {\it defects} of
the normal random walk state taken as a reference. Expressing the random
walk hypothesis as the fully symmetric state of the market, we show that
trading patterns corresponds to local breakdowns of this symmetry. Based on
simple symmetry principles, we show how to classify the possible patterns
based on the measurements of three fundamental dimensionless numbers that
are found to characterize a given market regime. The first one is the
Froude number
defined by
\begin{equation}
F\equiv {\frac{v^2}{p_0g}}~,
\end{equation}
where $p_0$ is the current price level, $v$ (resp. $g$) is the price
velocity (resp. acceleration). It measures the relative strength of the
trend with respect to the acceleration/deceleration. The second
dimensionless number is the future over past horizon ratio, defined as the
ratio
of the forecast time over the time interval used for detecting the pattern.
The third relevant parameter is the dimensionless time
interval $T_N$ used for detecting the pattern, i.e. the
reduced dimensionless learning period, expressed as follows,
$$
T_N=\frac v{p_0}t_N,
$$
where $t_N$ is the learning period, or past horizon, expressed in some
dimensional time units (days for instance).

\section{Fundamental symmetry principles}

Consider a time series for the price $S$ of an asset. It is defined as a
sequence of $N+1$ values of the price $S(t_0), S(t_1),..., S(t_N)$, given
for $N+1$ equidistant successive moments of time $t=t_j,$ where $j=0,1,2...N$%
. Our aim is to unravel indicators, i.e. departure from randomness, for the
level of the price $S$($t_N+\Delta t$) at a later time $t=t_N+\Delta t$,
that are compatible with the following five general properties.

\begin{itemize}
\item  Prices should remain positive, $S>0$, and their theoretical
description should be invariant with respect to a change of units: the
number describing the value of a stock changes when expressing it in US
dollar or in Euros, but the stock value remains what it is. The detection of
a price pattern and a forecast must thus remain invariant with respect to
such changes. Within the efficient market hypothesis, this requirement
eliminates Bachelier's model of the random walk of prices and replaces it by
the random walk of the logarithm of prices \cite{Samuelson}. The model of a
random walk of the logarithm of the price has, by definition, the symmetry
of translational invariance of returns, which in turn is equivalent to the
symmetry of scaling invariance of the price: the multiplication of all
prices by the same factor does not change their return. Such scale
invariance symmetry underlies many natural phenomena and is a strong
constraint for theoretical construction \cite{Dubrulleetal}.

\item  The theory should be invariant with respect to a change of time unit.

\item  A sensible theory of prices should not lead to zeros, poles or
divergences in finite time, as prices and values are finite.

\item  Patterns and their associated forecast should be defined in
probabilistic terms, allowing for multiple scenarii evolving from the same
past evolution; probabilities for different scenarii should be expressed
only through historical data, and in the form constrained by the above
mentioned demands on the price. Deeply imbedded in our approach is the view
of the future as a set of potentially acceptable trajectories that can
branch and bifurcate at special times. At certain times, only one main
trajectory extrapolates with high probability from the past making the
future depend almost deterministically (albeit possibly in a nonlinear and
chaotic manner) on the past. At other times, the future is much less certain
with multiple almost equivalent choices. In this case, we return to an
almost random walk picture. The existence of a unique future must not be
taken as the signature of a single dynamical system but as the collapse of
the large distribution of probabilities. This is the concept learned for
instance from the famous Polya Urn problem in which the historical
trajectory appears to converge to a certain outcome, which is however solely
controlled by the accumulation of purely random choices; a different outcome
might have been selected by history with equal probability \cite{PolyaUrn}.
We propose that it is fundamental to view any forecasting program as
essentially a quantification of probabilities for possible competing
scenerios. This view has been vividly emphasized by Asimov in his famous
Science-Fiction ``Foundation'' series \cite{Asimov}.

\item  We restrict the theoretical formulation to the case where, in absence
of the fundamental equations for price dynamics and other knowledge, the
future values of the price are constrained only by the past values. In order
to reduce substantially the class of possible scenarii, we propose to
express this condition in a way that makes apparent a deep symmetry between
time evolution and functional mapping. We will explain below how this
theoretical program can be formalized by the symmetry of {\it functional
self-similarity} \cite{YG}.
\end{itemize}

Our approach can be thought of as a search for a non-autonomous dynamical
system, where the law of motion changes with time and its functional form
remains unknown. Such a dynamical system is not self-similar in real time,
does not have a single dynamical representation, but can be characterized by
a functional self-similarity \cite{YG}. We do not assume that the basic laws
which govern market evolution should remain the same in the future as they
were in the past, although they will be obtained from the system's past. In
our approach, they are prescribed to evolve with time, due to the
assimilation of new information about the market. One can also think about a
window of forecasting detected in the market evolution as a {\it spontaneous}
breaking of continuous translational time invariance (the random walk
reference being translationally time invariant in its increments), occurring
whenever it is dictated by relative probabilities of the evolution patterns
with and without explicit violation of this symmetry.

In addition to these five natural requested properties, we add the concept
that

\begin{itemize}
\item  deviations from the random walk hypothesis are quantified by
measurements of the price velocity $v$ and of the price acceleration $g$.
\end{itemize}

This additional ingredient is motivated by the following considerations.

\begin{enumerate}
\item  First, as we recalled in the introduction, most technical trading
systems attempt to measure trends and/or change of trends in one way or
another. We thus adopt the pragmatic point of view that decades of empirical
research by practitioners has unraveled indicators of potential value that
may capture useful information.

\item  A complementary view point is that the market is self-organized by
the action of all traders, many of whom use technical analysis to guide
their investment decisions. It thus makes good sense to use indicators that
are a decisive part of the creation of the very structure of the market that
one tries to detect and from which one would like to forecast. This is in
spirit similar to the approach advocated by models of the stock markets
viewed as complex self-organized adaptive systems \cite{Santefe}.

\item  Another argument is that a constant trend in the logarithm of the
price simply defines a fixed return rate, which in economic theory can be
interpreted as the risk-free interest rate plus the risk premium paid for
being invested in the market. The velocity is thus similar to a
risk-adjusted return, a fundamental quantity in Portfolio theory and
practice. We propose the concept of a ``psychological Galilean principle'',
according to which investors perceive so-called ``market forces'' only when
trends change. We borrow here on the physiological and psychological
evidences \cite{psychology} that a constant stimulus is progressively
endogeneized and decreases progressively from the conscious mind. A
variation (acceleration/deceleration) is needed to create a new stimulus.
Similarly, we argue that investors are more sensitive, after a while, to
change of trends rather than to the continuation of the trend.
\end{enumerate}

\section{Scenarios and Probabilities}

\subsection{Velocity-acceleration parameterization}

The first step of our theoretical construction consists in selecting a
parametric representation of the past time series. As we argued above, the
information of the past price realizations ending at any given time $t_N$
(called ``the present'') is encoded by two parameters, the price velocity $%
v(t_N)$ and the acceleration $g(t_N)$. The simplest non-trivial use of this
parameterization is to form the second-order regression polynomial,
\begin{equation}
\label{hahbaj}S_0(t)=A_0+A_1 t+A_2 t^2,
\end{equation}
where $S_0(t)$ is the model price at time $t_0\leq t\leq t_N$, with
coefficients $A_0,$ $A_1,A_2$ adjusted to the real time series $S(t)$ by a
mean-square fit or any other suitable regression technique. The coefficients
$A_0,$ $A_1,A_2$ are obtained as the solutions of a system of a linear
algebraic equations derived from the condition of minimal Euclidean distance
between the polynomial (\ref{hahbaj}) and historical prices. It is clear
that $A_1$ (resp. $A_2$) is proportional to the price velocity (resp.
acceleration). This representation (\ref{hahbaj}) is reminiscent of the
so-called ``parabolic curve pattern'' often used in technical analysis (see
for instance $http://www.chartpattern.com/paraboliccurve.html$). The
parabolic formula follows from an implicit Newtonian dynamics. Analogies
with classical mechanics are deeply ingrained within both modern economics
and technical analysis. Absence of fundamental equations makes the task of
building any statistical or quantum mechanics of market process extremely
difficult, if possible. Nevertheless, based on symmetry, we are able to
formulate probabilistic market dynamics, or a Gibbsian-like statistical
market mechanics, by starting from this Newtonian-like representation (\ref
{hahbaj}). As already mentioned, the expression (\ref{hahbaj}) is a
convenient point of departure but other parameterizations are possible as
long as they capture the two fundamental quantities $v$ and $g$.

The representation of the price sequence $S(t_0),S(t_1),...,S(t_N)$ by (\ref
{hahbaj}) filters out the high-frequency variations of the price around the
trend and its variation. This natural filter is usually performed to get rid
as much as possible of the noise decorating such trend and acceleration. It
is important to stress that the real price quotes $S(t_i)$ carry a lot of
``noise'' and not only information. Using a mean-square fit assumes that the
residuals are Gaussian noise and implies that the coefficients $A_0,$ $%
A_1,A_2$ are linearly dependent on the quoted prices $%
S(t_0),S(t_1),...,S(t_N)$. As we will show, this guarantees scaling
invariance of the theory. For negative accelerations $A_2<0$, the model
price $S_0(t)$ given by (\ref{hahbaj}) is {\it not positively defined} for
arbitrarily large future times, in contradiction with our above requirement.
It is thus forbidden to use it directly for an extrapolation. However, we
will show how it can be exploited extensively as a source of both
qualitative and quantitative information about the future evolution of the
market. In fact, the ``bare'' model price $S_0(t)$ can be used for qualitative
predictions of the direction of the price movements, leaving aside magnitude
of the moves.

We rewrite $S_0$ using a dimensionless time, so that it becomes explicit
that $S_0$ remains invariant with respect to change of time units
\begin{equation}
\label{nhakak} S_0(T,F)=A_0(1+T+F^{-1}T^2)~,
\end{equation}
where $T$ is the dimensionless ``reduced'' time,
\begin{equation}
\label{habhan}T=\frac{A_1}{A_0}t~,
\end{equation}
whose sign follows that of the velocity $A_1$.
The expression (\ref{nhakak}) is invariant with respect to a change of time
units, since the coefficients $A_p$ are transformed into $k^{-p}~A_p$
under the transformation $t \to k t$. Due to this invariance,
it is convenient to think of the
beginning of the time series $T_0$ as the origin of time $T_0=0$ and take
the last known time (present) $T_N$ accordingly. The dimensionless Froude
number
$F$ measures the relative strength of the price velocity $A_1$ compared to
the price acceleration $A_2$, given the price level $A_0$ attained at time $%
0 $:
\begin{equation}
\label{Froude}F=\frac{A_1^2}{A_2~A_0}~.
\end{equation}
The sign of $F$ is determined uniquely by that of the acceleration $A_2$.
This dimensionless ratio (\ref{Froude}) is well-known in hydrodynamics as
the {\it Froude} number. In the hydrodynamical context, this dimensionless
number relates the ratio of inertia to buoyancy forces and is applicable in
particular to homogeneous shallow water flow, or two layers flow.
Explicitly, in the shallow water approximation, the Froude number is $%
F=U^2/(gH)$, in which $U$ is the characteristic velocity, $H$ the
characteristic fluid depth and $g$ the acceleration due to gravity. Rather
than an analogy with hydrodynamic turbulence \cite{analturb} controlled by
the Reynolds number weighting the nonlinear convective forces against the
viscous forces, this analogy with gravity-controlled viscousless fluids is
very suggestive: for viscousless liquids flowing above obstacles, two
regimes can occur. (i) For Froude numbers remaining always less than one,
the flow is only weakly perturbed by the obstacle. (ii) When the Froude
number reaches one at some point above the obstacle, a major perturbation of
the flow appears with a so-called hydraulic jump \cite{hydrojump}. We will
show below that similar regime transitions occur in the dynamics of market
price as a function of the Froude number (\ref{Froude}). The phenomenology
of the market patterns is however significantly richer due to the importance
of the signs of both $A_1$ and $A_2$.

\subsection{Functional Self-similarity and probabilistic scenarii}

For each present time, we obtain a parabolic parameterization (\ref{nhakak})
that provides a robust coarse-grained representation of the information of
past time prices. In principle, we may use much higher order polynomials or
different nonlinear functions to represent more accurately the many degrees
of freedom of the market price dynamics and then use these representation to
extrapolate into the future. After initial optimism, this dynamical system
approach \cite{Farmer2,Lebaron} does not live up to expectations \cite{Peters}.
In addition, there are important technical problems in specifying
high-order nonlinearities from noisy stock market data \cite{Farmer2}.

In contrast, we view the parabolic parameterization (\ref{nhakak}) as the
unique projection of a large set of possible equivalent trajectories over
the learning time interval $0\leq T\leq T_N$. Only one of them will be
selected by the dynamical evolution in the future. In order to construct
this set of trajectories that allow for multiple scenarii, we view the
parabolic parameterization (\ref{nhakak}) as the lowest order expansion {\it
in the space of functions} containing more complicated functional forms,
i.e. scenarii for the future. Our approach amounts to map the extrapolation
in the future onto an evolution in the space of functions on increasingly
complex ``approximants'' (scenarii) $y_n(t)$, which are linked to each
other by the
symmetry of functional self-similarity. This approach is very natural since
the future evolution can always be encoded by some mathematical representation;
the challenge is then to guess how to restrict the set of such
representations to achieve predictability.

The symmetry of functional self-similarity allows us to reach this goal by the
condition that different functions (scenarii) are not completely independent
from each other but can be seen as constituting a hierarchical construction or
multishell structure, such that the successive layers (approximations)
$y_{n\text{ }}(t)$, for
$n=0,1,...,$ are viewed as ordered realizations of a dynamical system evolving
with respect to the approximation number $n$ playing the role of an
effective time in the
functional space. The symmetry of functional self-similarity is expressed
mathematically
by self-similar renormalization group equations acting on the
approximants:
\cite{YG}
\begin{equation}
\label{funciaooa}y_{n+m}(\varphi )=y_n(y_m(\varphi ))~,
\end{equation}
where all approximations are expressed as a function of a
zeroth-order-approximation, through the relationship $y_0(t)\equiv \varphi$.
In words, this property (\ref{funciaooa}) of functional self-similarity
means that the same functional relationship relates the approximant of order
$n-1$ to the approximant of order $n$ as the approximant of order $n$ to the
approximant of order $n+1$. The relation (\ref{funciaooa}) expresses that
all approximations are connected by an identical iteration procedure of
successive improved approximations. This property also ensures the fastest
convergence criterion \cite{YG}, i.e. the strongest stability for the
selected scenarii. Going from the abstract space of approximations to the
real time, we thus obtain a set of self-similar extrapolation functions,
corresponding to the different possible future market dynamics.

Using the dimensionless variables $T$ and $F$ and following the general
procedure of algebraic self-similar bootstrap \cite{YG} recalled in the
appendix, we deduce from the
polynomial (\ref{nhakak}) the three simplest polynomial approximations
\begin{equation}
S_{00}=A_0,\quad S_{01}=A_0(1+T),\quad S_{02}=A_0(1+T+F^{-1}T^2)~.
\end{equation}
They correspond to successive truncations of (\ref{nhakak}) at increasing
orders of the power
of $T$. Implementing the algebraic self-similar bootstrap
procedure \cite{YG} recalled in the appendix, we obtain two non-trivial
approximants, which give two
possible scenarii for the future price evolution:
\begin{equation}
\label{sen1}S_1(T)=A_0\exp (T)~,
\end{equation}
\begin{equation}
\label{sen2}S_2(T,F)=A_0\exp \left( T\exp \left( \frac TF\right) \right) ~.
\end{equation}
Since $T$ and $F$ can be of both signs, $S_1(T)$ can be increasing or
decreasing and $S_2(T)$ can be in addition non-monotonous.

These two scenarii-approximants (\ref{sen1},\ref{sen2}) satisfy all symmetry
demands formulated in section II. The approximant $S_1$ given by (\ref{sen1}%
) is quite special since it possesses the self-similarity symmetry both in
real time $T$ and in the space of approximations. Indeed, a time translation
leads solely to a redefinition of $A_0$, while keeping exactly the same
{\it functional} form. In contrast, the approximant $S_2$ is not invariant
under a
time translation and possesses only the functional self-similarity (\ref
{funciaooa}). It thus corresponds to a breaking of the time-translation
symmetry.

The two scenarii-approximants (\ref{sen1},\ref{sen2}) corresponds to two
different forecasts for the future price evolution. What is the probability
of each scenario? We follow ref.\cite{multiGY} and assume that the most
probable scenario corresponds to the most stable approximant with respect to
a change of the parabolic parameterization (\ref{nhakak}). In other words,
the stability of each scenario is estimated by calculating the amplitude of
its variation upon a change of the parameters of the parabolic parameterization
(\ref{nhakak}). This allows us to use the concept of a Lyapunov exponent in
the space of approximants. Then, using ideas from
dynamical system theory, we generalize the dynamical
Kolmogorov-Sinai entropy for the finite time behavior and stable as well as
unstable trajectories. In our present context, the generalized
Kolmogorov-Sinai entropy of an approximant is nothing else but the Lyapunov
exponent associated to a variation of parabolic parameterization
(\ref{nhakak}).
As in \cite{multiGY}, statistical physics then teaches us that a
probability is obtained by taking the exponential $p\sim e^{-S}$ of minus
the entropy and then normalizing to one. Since the generalized
Kolmogorov-Sinai entropy is nothing else but the entropy rate, it defines
the entropy. Movement along a stable trajectory decreases the entropy
counted from the entropy of the initial state, while the movement along an
unstable trajectory increases the entropy. Probabilities are most
conveniently expressed through the so-called multipliers, defined as the
exponential of the Lyapunov exponents, and given by the functional
derivative
\begin{equation}
m_k=\frac{\delta \ S_k}{\delta \ S_1},\ k=1,2~.
\end{equation}
which yields
\begin{equation}
m_1(T)=1;\quad \quad m_2(T,F)=\left( 1+\frac TF\right) \exp \left[ T\ \left(
\frac 1F-1+\exp \left( \frac TF\right) \right) \right] ~.
\end{equation}
The corresponding probabilities $p_1$ and $p_2$ for each scenario are
inversely proportional to the multipliers and in proper normalization can be
written as \cite{multiGY}
\begin{equation}
p_1(T,F)=\frac 1{1+\left| m_2(T,F)\right| ^{-1}},\quad p_2(T,F)=\frac{\left|
m_2(T,F)\right| ^{-1}}{1+\left| m_2(T,F)\right| ^{-1}}~.
\end{equation}
We define the average of the two scenerios as
\begin{equation}
\label{jakjlklal}S^{*}(T,F)=p_1(T,F)\ S_1(T)+p_2(T,F)\ S_2(T,F)~.
\end{equation}

Probabilities defined in this way put more weight onto the trajectories with
small multipliers which are the most stable, while not forbidding completely
unstable trajectories with multipliers larger than one. One can imagine
situations where all trajectories are unstable or neutral, with multiplier
equal to or larger than 1. Then, the least unstable trajectory will receive
more weight
than the more unstable ones.

\section{Classification of market phases}

The classification of market patterns
depends on three parameters:
\begin{itemize}
\item the dimensionless time horizon $\Delta T/T_N$ normalized by the time
interval $T_N$
over which the parabolic representation is constructed,
\item the reduced dimensionless learning period $T_N$ itself,
\item the Froude number $F$.
\end{itemize}

In order to classify the different possible temporal patterns of market
prices, one should in principle compare the forecasted value
$S^{*}(T_N+\Delta T,F)$ to
the ``present'' price approximated by $S_0(T_N,F)$, where $T_N$ and $\Delta
T $ are the dimensionless times obtained from $t_N$ and $\Delta t$ by the
transformation (\ref{habhan}). We propose to use the average scenario
$S^{*}(T,F)$
instead of $S_0(T_N,F)$ in order to reduce or eliminate as much as possible
any
systematic errors. Indeed, the use of the same function $S^{*}(T,F)$ provides
a scheme for cancellation of errors that would not otherwise occur if the
the parabola $S_0(T_N)$ given by (\ref{hahbaj}) was chosen instead.

A given market pattern is thus determined by the behavior fo the predicted
return
\begin{equation}
R=~\ln \left( {\frac{S^{*}(T_N+\Delta T,F)}{S^{*}(T_N,F)}}\right) ~.
\label{fnalnacl}
\end{equation}
We classify the different regimes by looking at the sign of the
return and the transition between two regimes is quantified by the condition
$R=0$ which is equivalent to
\begin{equation}
\label{quyqjq}\Delta S^{*}=S^{*}(T_N+\Delta T,F)-S^{*}(T_N,F)=0~.
\label{condjuak}
\end{equation}
The number of solutions of equation (\ref{quyqjq}) can be $0,1,2$ or $3$.

The following four combinations of the parameters are
possible and define the following four regimes.

\begin{enumerate}

\item  ``Super-bull'' ($\Delta T>0,\ F>0$) corresponding to positive price
velocity and positive acceleration; we depict this regime with the following
pictograph $\rfloor$.

\item  ``balanced-bull'' ($\Delta T>0,\ F<0$) corresponding to positive price
velocity and negative acceleration; we depict this regime with the following
pictograph $\lceil$.

\item  ``Super-bear'' ($\Delta T<0,\ F<0$) corresponding to negative price
velocity and negative acceleration; we depict this regime with the following
pictograph $\rceil$.

\item  ``Balanced-bear'' ($\Delta T<0,\ F>0$) corresponding to negative price
velocity and positive acceleration; we depict this regime with the following
pictograph $\lfloor$.

\end{enumerate}

In each case, one will find qualitatively different diagrams for the returns.
A ``Bull'' regime of evolution corresponding to $R>0$ will intermingle with a
``Bear'' regime corresponding to $R<0$ in the phase space of control
parameters in a way peculiar to each of the four cases.

Figure \ref{fig1}-\ref{fig6} presents some general statistics obtained from
this
classification. We have analyzed daily returns of twelve data sets
comprising some major indices
\begin{itemize}
\item the Dow Jones index from Jan. 2, 1970 till Feb. 24, 1998,
\item the SP500 index from Jan. 1, 1950 till June 1, 1999,
\item the Nasdaq index from Feb. 5, 1971 till May 18, 1999,
\item the German DAX index from July 1, 1991 till May 17, 1999,
\item the British FTSE index from April 17, 1990 till May 17, 1999, and
\item the Japanese Nikkei index from April 16, 1990 till May 17, 1999,
\end{itemize}
some major bonds
\begin{itemize}
\item the thirty year US treasury bond TYX from Oct. 29, 1993 till  Aug. 9,
1999,
\item the Japanese Governement Bond JGB from from Jan. 1, 1992 till March
23, 1999
\end{itemize}
and some major currencies against the US dollar all from Jan. 4, 1971 till
May 19, 1999,
\begin{itemize}
\item the British pound GBP,
\item the Swiss franc,
\item the German mark,
\item the Japanese Yen.
\end{itemize}

Each figure in the series \ref{fig1}-\ref{fig6} presents first the time
evolution of the price (top). The middle plots represents the Froude number
defined
in equation (\ref{Froude}) as a function of the reduced prediction horizon
$\Delta T \equiv (A_1/A_0) \delta t$, where $\delta t$ is fixed
equal to 5 days. The four quadrants sampled clockwise correspond to the four
regimes $\rfloor$, $\lceil$, $\rceil$ and $\lfloor$ defined above. The
bottom plots
quantify the relative frequency of each regime. Specifically, for each
regime, we
look at the sign of the return prediction and count the number of time
a given regime with a given return sign has been predicted. In this way, we
define
eight patterns of which six fundamental ones remain to be considered by
looking
successively clockwise at the quadrants of the diagram of the Froude as a
function of $\delta T$:
\begin{itemize}
\item $p_1 \equiv \rfloor +$: super bull predicting a positive return,
\item $p_2 \equiv \rfloor -$: super bull predicting a negative return
(impossible)
\item $p_3 \equiv \lceil +$: balanced-bull predicting a positive return,
\item $p_4 \equiv \lceil -$: balanced-bull predicting a negative return,
\item $p_5 \equiv \rceil +$: super bear predicting a positive return
(impossible)
\item $p_6 \equiv \rceil -$: super bear predicting a negative return,
\item $p_7 \equiv \lfloor +$: balanced-bear predicting a positive return,
\item $p_8 \equiv \lfloor -$: balanced-bear predicting a negative return.
\end{itemize}
Note that $p_1 \equiv \rfloor +$, $p_3 \equiv \lceil +$, $p_6 \equiv \rceil
-$ and
$p_8 \equiv \lfloor -$ are trend-following patterns while $p_2 \equiv
\rfloor -$,
$p_4 \equiv \lceil -$, $p_5 \equiv \rceil +$ and $p_7 \equiv \lfloor +$ are
contrarian patterns. Of these four contrarian patterns, only
$p_4 \equiv \lceil -$ and $p_7 \equiv \lfloor +$ are allowed within the
super-exponential
framework. The theory thus accounts for a natural preferential bias in
favor of
trend-following patterns, while contrarian patterns do appear with
non-negligible frequency.

The three dotted, dashed and continuous lines represent the transition between
different signs of predicted returns from the condition (\ref{condjuak}):
above (resp. below) the dotted line in the second
(resp. third) quadrant, the predicted return is negative. It then turns
positive between the
dotted and dashed line, then negative again between the dashed and the
continuous line and
positive below (resp. above) the continous line in the second (resp. third)
quadrant.

Since market noise is
an important issue, we also investigate how robust is a given pattern with
respect to the amplitude of the predicted return: we count only those patterns
with a predicted amplitude of the price variations
$\Delta S^{*}$ defined by (\ref{quyqjq}) larger than a threshold defined as
a multiple
of the standard deviation of the fit of the price by the parabola
$S_0(T_N)$ given by (\ref{hahbaj}) in the training window of length $t_N$.
The result is presented in the plots at the bottom of figures
\ref{fig1}-\ref{fig6}
showing the number of realizations of each of the six relevant patterns as
a function of
the threshold. When the threshold increases, some predictions are left over
since
their predicted price variation is below the threshold: we thus expect
that the number of realizations of each pattern should decays as the
threshold increases.
We observe that the relative frequency of a given pattern decays typically
exponentially with the threshold, in agreement with the approximate exponential
character of the distribution of daily price variations \cite{lahesor} and
of drawdowns
\cite{outlier1,outlier2}. Furthermore, among all patterns, the
balanced-bull predicting negative returns ($p_4 \equiv \lceil -$) and
balanced-bear predicting positive returns occur less frequently than the other
patterns, showing a preponderance for trend-following patterns.

\subsection{Super-bull $\rfloor$ ($\Delta T>0,\ F>0$)}

We start by a word of caution as this scenario is rather unstable. Indeed,
In this case of positive dimensionless time and Froude number,
the multiplier $m_2(T,F)$ is always
larger than one, which means that the second scenario $S_2(T,F)$ is always
unstable. When the Froude number decreases, both $m_2(T,F)$ and $S_2(T,F)$
grow larger and larger, becoming very large for sufficiently small
$F$. However, the mathematical divergence only occurs at infinite times.
Strickly
speaking, this does not violate the principle requested by our theory that
 price remaining finite; however, in practice, it may lead to instabilities.
 The formal definition of the finite
average $S^{*}$ still holds mathematically speaking but the
accuracy may be problematic. Limiting the theory to a second-order regression
(i.e. solely in terms of the velocity and the acceleration)
is a limiting feature of our present approach and should be improved by
considering higher-order approximants.

In the super-bull case, the prediction is that returns are
expected to be always positive corresponding to pattern $p_1 \equiv \rfloor +$.

Figure \ref{figsupbull1},\ref{figsupbull2} presents statistical tests of this
prediction on the twelve data sets shown in figures \ref{fig1}-\ref{fig6}.
In these
plots, the learning interval is fixed to 15 days and the prediction horizon is
fixed to 5 days. For each
asset, two plots are presented. The top one shows the normalized number of
successes
for the prediction of the sign of the return as a function of threshold,
where the threshold
is defined as in figures \ref{fig1}-\ref{fig6}. A normalized success close to
$0.5$ corresponds to a $50\%$ probability of being right or wrong and is
thus undistinguishable
from chance. The issue however is more subtle because the existence of
trends in
certain markets such as in indices bias this estimation. To account for
such bias and
others stemming from specific structures of the distributions of daily
returns in the raw
time series, we have generated 1000 surrogate times series for each of the
12 assets by
reshuffling at random the daily returns. On each of the 1000 surrogate time
series, we
have applied our procedure and have measured the success rate as a function
of threshold,
following the same methodology
as for each initial time series. This allows us to plot the $90\%$
(continuous line) and $99\%$ (dotted line) confidence levels, defined by
the fact that
900 (resp. 990) among the 1000 surrogate times series gave a
success rate in the interior of the domain bounded by the
continuous (resp. dotted) line.

In addition, since our framework incorporate both a trend-following component
(quantified by the velocity) and contrarian ingredient (quantified by the
acceleration), it is instructive to compare its performance with two
strategies.
\begin{enumerate}
\item The first one is a simple trend-following strategy and consists of
taking the linear approximation $A_0 (1+T) = A_0 + A_1 t$ of the
simple exponential approximant $S_1$ as the best prediction. This strategy is
independent of
the acceleration $A_2$ and is thus a pure trend-following strategy: it rises
(resp. decreases) if
the velocity $A_1$ is positive (resp. negative).
\item The second strategy consists in using the bare
parabolic parameterization (\ref{nhakak}), which represents the best
representation
of the local market behavior incorporating the interplay between trend and
acceleration without any theoretical improvement.
\end{enumerate}
In this way, we can really quantify the value, if any, brought by our
theory, compared
to more traditional technical analysis methods.

The bottom plot for each asset in figure \ref{figsupbull1},\ref{figsupbull2}
gives the normalized number of times the pattern
has been found as a function of the threshold, this for the three different
strategies,
namely
our best prediction based on the average $S^*$ over the two approximants, the
trend-following strategy (linear approximation of $S_1$) and the parabolic
parameterization
(\ref{nhakak}).
This plot is important in order to assess the quantitative importance of a
given succes rate
in terms of its frequency. Consider for instance the Swiss franc. We
observe that both the
trend and average approximant exhibit statistically significant predictive
power with
two notable peaks as a function of threshold, approximately at $0.5$ and $1.2$,
with a success rate approaching $60\%$ in the first case and overpassing
$80\%$ in the second case. The apparent overwhelming superiority of the
second  peak
is moderated by the fact that it concerns about 1/10th of the cases covered
by the
first peak for the average approximant and about 1/100th of the case
covered by the first peak for the trend-following strategy. Their impact for
a successive investment strategy will thus be an interplay between their
success rate
and the occurrence rate.

The following overall picture emerges from examination of the twelve markets
shown in figure \ref{figsupbull1},\ref{figsupbull2}.
Taken together, we find that the trend-following and the average approximant
strategies exhibit statistically significant success rates, while the parabolic
strategy is not different from random coin tossing. While the trend-following
strategy seems to exhibit sometimes a better performance than the average
approximant,
it is much less robust in terms of its number of occurences.

\subsection{Balanced-Bull market $\lceil$ ($\Delta T>0,\ F<0$)}

In the balanced-bull regime, both positive and negative returns can be
predicted.
A ``phase diagram'', defined in the parameter space $(\Delta T/T_N, |F|)$ with
$T_N = 1$ and shown in figure \ref{FigBalbull}, summarizes the different
possible cases.
>From the condition
of positive prices $S_0>0$, we find that $F$ must satisfy the condition
$$
|F|>F_0=\frac{T_N^2}{1+T_N}\ .
$$
There are three branches that solve equation (\ref{quyqjq}): $F_1(\Delta T)$,
$F_2(\Delta T)$ and $F_3(\Delta T)$, with $\left| F_1(\Delta T)\right|
<\left| F_2(\Delta T)\right| <\left| F_3(\Delta T)\right| $. These curves
are shown in  figure \ref{FigBalbull} as dotted-dashed, dashed and
continuous lines
respectively. Below the curve
$\left| F_1(\Delta T)\right|$, $R$ is always negative, corresponding to a
trend-reversing forecast for all possible time horizons, while $R$ becomes
positive above the curve, corresponding to the trend-following forecast for
large $\left| F\right|$. There is also a ``tongue''-shaped region of
trend-reversing regime,
encircled by the curves $\left| F_2(\Delta T)\right| $ and $\left|
F_3(\Delta T)\right|$. We are inclined to interpret this region as an
artifact appearing as a remnant of the non-renormalized phase-equilibria
curve.

It is instructive to calculate the line delineating the change of sign of
the return,
which would follow directly from the non-renormalized price
$S_0$, given by the condition ${S_0(T_N+\Delta T,F)-{S_0(T_N,F)=0,}}$ and
compare it to the phase-equilibria for the renormalized price. This line is
shown in figure \ref{FigBalbull} as the long-dashed line and is refered to
as ``regression''.

These results can be understood intuitively as follows. A
small absolute value of the Froude number implies a large (negative)
acceleration, hence the possibility for the price to change course over
the prediction horizon. We thus expect and observe that the predicted return
is negative {\it below} a transition line, both for the non-renormalized price
parabola $S_0$ and for our prediction $S^*$. We observe that the effect of the
functional renormalization is to shift significantly the transition line
towards lower $|F|$, i.e. larger negative accelerations. This can be seen
to result in part from the stabilization effect of the positivity property of
super-exponentials.

We note that these results are robust with respect to a change of the
dimensionless learning period
$T_N$, which has only a marginal influence only on the quantitative shape
of the phase
diagram but not on its qualitative properties. In fact, the topological
properties of the transition lines never change when varying $T_N$.

Figure \ref{FigBalbullret} shows the predicted return defined by equation
(\ref{fnalnacl}) as a function of the ratio $\Delta T/ T_N$ of the
prediction horizon
over the learning interval for several values of the Froude number. The
three boundaries
$F_1(\Delta T)$, $F_2(\Delta T)$ and $F_3(\Delta T)$ shown in
figure \ref{FigBalbull} can be deduced qualitatively from
this figure \ref{FigBalbullret} from the three regimes where $R$ is always
positive
(large $|F|$),$R$ changes signs with a cusp (intermediate $|F|$) and is always
negative (small $|F|$).

These predictions are tested in figure
\ref{figp3p4testa},\ref{figp3p4testb},\ref{figp3p4testc},\ref{figp3p4testd}
which
presents statistical tests on the twelve data sets shown in figures
\ref{fig1}-\ref{fig6}.
The same parameters as for the super bull regime tested in figure
\ref{figsupbull1},\ref{figsupbull2}
have been used
(the learning interval is fixed to 15 days and the prediction horizon is
fixed to 5 days).

Pattern 3 corresponding to the trend-following pattern $\lceil +$ is first
shown.
We observe that the parabolic prediction (open circles) and the average
approximant
(crosses) strategies exhibit statistically significant success rates, while
the
trend-following
strategy is not different from random coin tossing. While the parabolic
prediction
strategy seems to exhibit sometimes a better performance than the average
approximant,
it is much less robust in terms of its number of occurences.

For pattern 4 corresponding to the contrarian pattern $\lceil -$,
we show only two strategies, the average approximant (crosses) and the
bare parabolic parameterization (\ref{nhakak}) represented by open circles,
since the trend-following strategy never predicts the pattern $\lceil -$ by
definition. The average approximant clearly exhibits
statistically significant success rates for most of the assets, while the
parabolic strategy is not different from random coin tossing in this
explored threshold range.

\subsection{Super-bear $\rceil$ ($\Delta T<0, F<0 $)}

We start by a word of caution as this scenario is rather unstable.
In this case of negative dimensionless time and Froude number,
the multiplier $m_2(T,F)$ and its associated scenario $S_2(T,F)$
both go to very small values with decreasing $\left| F\right|$ in the limit
of small $\left| F\right|$. The average approximant $S^{*}$
goes to zero as well. The large variations are symptomatic of potential
instabilities. Limiting the theory to a second-order regression
(i.e. solely in terms of the velocity and the acceleration)
is a limiting feature of our present approach for the super-bear
(as it was the case for the super-bear discussed above) and should be
improved by
considering higher-order approximants.

In the super-bear case, the prediction is that returns are
expected to be always negative corresponding to pattern $p_6 \equiv \rceil -$.
Pattern $p_7 \equiv \rceil +$ is thus impossible within the present framework
limited to a characterization of the market price evolution solely in terms
of a velocity and an acceleration.

Figure \ref{figsupbear1},\ref{figsupbear2} presents the same statistical
tests as in figure
\ref{figsupbull1},\ref{figsupbull2} of this
prediction on the twelve data sets shown in figures \ref{fig1}-\ref{fig6}.
In these
plots, the learning interval is fixed to 15 days and the prediction horizon is
fixed to 5 days. As for the super-bull case, we observe that
the trend-following and the average approximant
strategies exhibit statistically significant success rates, while the parabolic
strategy is not different from random coin tossing. While the trend-following
strategy seems to exhibit sometimes a better performance than the average
approximant,
it is much less robust in terms of its number of occurences.

\subsection{Balanced-Bear $\lfloor$ ($\Delta T<0,\ F>0$)}

In the balanced-bear regime as for the previously
discussed balanced-bull regime, both positive and negative returns can be
predicted.
A ``phase diagram'', defined in the parameter space $(\Delta T/T_N, |F|)$ with
$T_N = 1$ and shown in figure \ref{FigBalbear}, summarizes the different
possible cases.

There are three branches that solve equation (\ref{quyqjq}): $F_1(\Delta T)$,
$F_2(\Delta T)$ and $F_3(\Delta T),$ with $F_1(\Delta T)<F_2(\Delta
T)<F_3(\Delta T)$. The phase diagram shown in figure \ref{FigBalbear} looks
very
similar to the balanced-bull case, with the region of positive
(resp. negative) returns below (resp. above)
the $F_1(\Delta T)$ curve. There is again a ``tongue''-shaped
region of trend-reversing regime,
encircled by the curves $F_2(\Delta T)$ and $F_3(\Delta T)$.
We propose to interpret this region as an
artifact appearing as a remnant of the non-renormalized phase-equilibria
curve. We show in addition the line of change of sign for the return
predicted by
the non-renormalized price $S_0$, given by the
condition ${S_0(T_N+\Delta T,F)-{S_0(T_N,F)=0,}}$.

For small $F$, the average approximant $S^*$ as well as the
non-renormalized parabolic
parameterization $S_0$ both predict a trend-reversing regime. The
interpretation is
the following: a small Froude number corresponds to a large positive
acceleration,
which has thus the capacity of reversing the negative trend over the
prediction horizon.
The main effect of our renormalization procedure is to shift the transition
line
downwards to smaller Froude numbers, i.e. to larger accelerations, a result
that
derives from the stabilization of our procedure.
This examplifies the non-trivial nature of the scenarii selected by the
self-similar functional renormalization group approach.
For larger $F$, i.e. smaller accelerations, the trend-following forecast
takes over.

The reduced dimensionless learning period $T_N$
is again a marginally relevant parameter, i.e. it influences the shape of
phase diagram only quantitatively. The shape and topology of
the phase equilibria curves are similar for all $T_N$.

It is interesting to notice that our classification of possible market
regimes exhibit a distinct asymmetry between bullish and bearish phases:
 a balanced-bull market and a balanced-bear
market will not evolve symmetrically in time. Most significantly, the
line of phase equilibria for the balanced-bear case saturates at large
$\Delta T/T_N$ at some constant value which is function only of $T_N$, while
for the balanced-bull case there is no such saturation.

Figure \ref{FigBalbearret} shows the predicted return $R$ defined by equation
(\ref{fnalnacl}) as a function of the ratio $\Delta T/ T_N$ of the
prediction horizon
over the learning interval for several values of the Froude number.
The three boundaries
$F_1(\Delta T)$, $F_2(\Delta T)$ and $F_3(\Delta T)$ shown in
figure \ref{FigBalbear} can be deduced qualitatively from
this figure \ref{FigBalbearret} from the three regimes where $R$ is always
positive
(small $F$),$R$ changes signs with a cusp (intermediate $F$) and is always
negative (large $F$).

These predictions are tested in figure
\ref{figp7p8testa},\ref{figp7p8testb},\ref{figp7p8testc},\ref{figp7p8testd}
which
presents statistical tests on the twelve data sets shown in figures
\ref{fig1}-\ref{fig6}.
The same parameters as for the super bull regime tested in figure
\ref{figsupbull1},\ref{figsupbull2}
have been used
(the learning interval is fixed to 15 days and the prediction horizon is
fixed to 5 days).

Pattern 7 corresponding to the trend-reversal or contrarian pattern
$\lfloor +$ is first shown.
Only two strategies are shown, namely the average approximant (crosses) and
the
bare parabolic parameterization (\ref{nhakak}) represented by open circles,
since the trend-following strategy never predicts the pattern $\lfloor +$ by
definition. The average approximant clearly exhibits
statistically significant success rates for most of the assets, while the
parabolic strategy is not different from random coin tossing.

For pattern 8 corresponding to the trend-following case $\lfloor -$,
we observe that the parabolic prediction (open circles) and the average
approximant
(crosses) strategies exhibit statistically significant success rates, while
the
trend-following
strategy is not different from random coin tossing. While the parabolic
prediction
strategy seems to exhibit sometimes a better performance than the average
approximant,
it is much less robust in terms of its number of occurences.

\section{Discussion and Conclusion}

In summary, we have presented a general framework that characterizes different
market phases viewed as local defects of the overall time-translational
invariance structure of the random walk reference. This framework is based
on the
price velocity and acceleration parameters and on a set of symmetry
principles, especially self-similarity of the prices and in the abstract
space of functional scenarii.

We have tested the quality of the predictions provided by our theory
by measuring the success rate of the prediction of the sign of the return
on twelve different assets and have compared the quality of these
predictions to those
from for traditional technical analysis including a trend-following and a
contrarian
strategy. The statistical significance of our results have been assessed by
generating
1000 surrogate time series for each of the twelve assets with exactly the same
statistical properties except for possible time-dependence, by reshuffling
the daily returns.
The application of the three
forecasting strategies to these 1000 surrogate times series shows that our
predictive
skill has overall a high degree of statistical significance. Furthermore,
we find that it
is robust over all phases of the market. This is in constrast with the
trend-following
strategy which is found to perform well only during
strong accelerating trends (categorized as ``super-bull'' and
``super-bear''). This is
also in contrast with the contrarian strategy which is found to perform
well only during
decelerating trends (categorized as ``balanced-bull'' and ``balanced-bear'').
Our probabilistic framework thus provides an automatic scheme for detecting
and selecting
what type of strategy is the best performer. This optimization relies on the
calculation of the stability of the different scenarii: the most stable one
is the most
probable and controls the strategy (trend-following or contrarian) that is best
adapted to a particular phase of the market.

The present framework can be seen as a generalization of the standard
``Newtonian'' deterministic technical analysis to a Gibbsian (statistical)
mechanics, in other words, to a probabilistic view of the future.
The appearance of probabilities has to be emphasized in constrast
to a deterministic view of future. This probabilistic framework captures
the inhomogeneity of market participants who can produce
 different forecasts based on equivalent information. It may thus
 provide a forum for reconciliation of the on-going feud between ``efficient
marketers'' and ``technicians'', by showing that the random walk paradigm and a
determinism view of the world
are different limiting cases of much broader and complex market
evolutions.

{\bf Acknowledgements}:  We are grateful to D. Darcet, A. Johansen
and V.I. Yukalov for stimulating discussions.

\pagebreak
\section*{Appendix: Statistical Self-Similar Analysis}

In this appendix, we recall the theory that leads to the results quoted
in section IIIB. We look at the prediction of the futur time evolution
of the market as a specification of
an a priori unknown function $f(x)$ of a real variable $x$ representing
the fundamental market trajectory as a function of physical time or the
dimensionless time.
We keep the notation $f(x)$ to be general.

The first step of the theory consists in recognizing that,
in the neighborhood of the present time $x=x_0$, one can define a series of
approximations $p_k(x,x_0)$ (also called asymptotic expansions) to this
function with
$k=0,1,2,\ldots$:
\begin{equation}
\label{1}
f(x)\simeq p_k(x,x_0), \qquad x\rightarrow x_0 ~.
\end{equation}
The number $k$ indexes the order of the approximations obtained according to
some construction scheme, which can be convergent
or divergent (i.e. asymptotic).

Following the algebraic self--similar renormalization procedure \cite{YG},
one introduces the algebraic transform defined by
\begin{equation}
\label{2}
P_k(x,s,x_0)=x^s p_k(x,x_0)\; ,
\end{equation}
where the exponent $s$ is yet unknown, and later will play the role of a
control
function. The inverse transform to that in Eq.(\ref{2}) is
\begin{equation}
\label{3}
p_k(x,x_0)=x^{-s} P_k (x,s,x_0)\; .
\end{equation}
The introduction of the transform (\ref{2}) in terms of power law products
keeps the symmetry requirements discussed in the main text and offers
an additional degree of freedom in the space of functions to impose
positivity and get
of rid of infinities (poles). Rather than constructing a trajectory in the
space of the initial approximations, the idea behind the introduction of
the transform (\ref{2}) is to deform smoothly the initial functional space of
the approximations $p_k(x,x_0)$ in order
to obtain a faster and better controlled convergence in the space of the
modified functions $P_k(x,s,x_0)$. This convergence can then be mapped back
to get the relevant estimations and predictions.

Technically, the procedure is as follows \cite{YG}. One first defines
an expansion function $x=x(\varphi,s,x_0)$ by the equation
\begin{equation}
\label{4}
P_0(x,s,x_0) = \varphi ,\qquad  x = x(\varphi,s,x_0) \; ,
\end{equation}
where $P_0$ is the first available term from the sequence $\{ P_k\}$.
This allows one to transform the problem of searching for an approximation
as a function of time into the problem of the dependence of the
 time as a function of the first approximant. This
type of transformation is well-known and a closely related one is for
instance used in hydrodynamics
under the name of the hodograph transform
to solve very non-linear flow problem \cite{hodograph} that are otherwise
out of reach
of standard functional methods. The variable $\varphi$
represents the running and continuous interpolation between successive
approximations in
the space of functions.

Once this change of variable is performed, all functions can be expressed
in terms of this new variable $\varphi$, which defines
\begin{equation}
\label{5}
y_k(\varphi,s,x_0) = P_k(x(\varphi,s,x_0),s,x_0)\; .
\end{equation}
The transformation inverse to Eq. (\ref{5}) reads
\begin{equation}
\label{6}
P_k(x,s,x_0)=y_k(P_0(x,s,x_0),s,x_0)\; .
\end{equation}
The family of endomorphisms, $\{y_k\}$, allows to define the ``velocity''
field
\begin{equation}
\label{7}
v_k(\varphi,s,x_0) = y_{k+1}(\varphi,s,x_0) - y_k(\varphi,s,x_0)~,
\end{equation}
corresponding to the change of the approximants per unit order of the
approximations.
The trajectory of the sequence $\{y_k\}$ is, by definitions (\ref{5}) and
(\ref{6}),
bijective to the approximation sequence $\{ P_k\}$.

>From the knowlege of the velocity as a function of the variable $\varphi$, one
defines an effective time increment $\tau$ for the evolution of the
dynamical system
in the space of functions: during that time increment $\tau$,
$P_k$ is transformed into $P_{k+1}$. This provides the evolution integral
\begin{equation}
\label{8}
\int_{P_k}^{P_{k+1}^*}\frac{d\varphi}{v_k(\varphi,s,x_0)} = \tau \; ,
\end{equation}
in which $P_k=P_k(x,s,x_0)$ is any given term from the approximation
sequence $\{ P_k\} ;\; P_{k+1}^*=P_{k+1}^*(x,s,\tau,x_0)$ is the best guess
obtained
in this approach for the fixed point of the
approximation series.  $\tau$ is an effective minimal time necessary
for reaching this fixed point.

Recall that we started with a sequence $\{ p_k\}$ of asymptotic
expansions of the function $f(x)$. We then passed to the
sequence $\{ P_k\}$ by means of the algebraic transformation (\ref{2}). We now
have to return back by employing the inverse transformation (\ref{3}). To
this end,
we set
\begin{equation}
\label{9}
F_k^*(x,s,\tau,x_0) = x^{-s} P_k^*(x,s,\tau,x_0)\; .
\end{equation}

The quantities $s$ and $\tau$ are the control functions guarantying the
stability of the method, that is, the convergence of the procedure. These
functions are to be defined by the stability conditions, such as the
minimum of multiplier moduli, together with additional constraints, like,
e.g., boundary conditions. Let us assume that we find from such conditions
$s=s_k$ and
$\tau =\tau_k$. Substituting these into Eq.({\ref{9}), we obtain the
self--similar
approximation
\begin{equation}
\label{10}
f_k^*(x,x_0) = F_k^*(x,s_k,\tau_k,x_0)\;
\end{equation}
for the function $f(x)$.

Let us now apply this theory to the polynomial perturbative series
\begin{equation}
\label{17}p_k(x)=\sum_{n=0}^ka_nx^n,\quad a_0\neq 0~,
\end{equation}
containing integer powers of $x$ as in (\ref{hahbaj}). Then, the
algebraic transform (\ref{2}) reads
\begin{equation}
\label{18}P_k(x,s)=\sum_{n=0}^ka_nx^{n+s} ~.
\end{equation}
The transform (\ref{18}) corresponds to an effective higher
perturbation order $k+s$, as compared to the initial series (\ref{17}) of
order
$k$. Eq.(\ref{4}) for the expansion function $x(\varphi,s)$ now reads
\begin{equation}
\label{19}P_0(x,s)=a_0x^s=\varphi\ ,
\end{equation}
which yields
\begin{equation}
\label{20}x(\varphi,s)=\left(\frac \varphi{a_0}\right)^{1/s}.
\end{equation}
The series of functions (\ref{5}) become
\begin{equation}
\label{21}y_k(\varphi,s)=\sum_{n=0}^ka_n \left(\frac
\varphi{a_0}\right)^{n/s+1}.
\end{equation}
The velocity field (\ref{7}) reads
\begin{equation}
\label{22}v_k(\varphi,s)=a_k\left(\frac \varphi{a_0}\right)^{1+k/s}.
\end{equation}
With the optimization condition $\tau =1$ corresponding to the
requirement that the fixed point is reached under a single iteration,
the evolution integral (\ref{8}) gives
\begin{equation}
\label{23}P_k^{*}(x,s)=P_{k-1}(x,s)\left[1-\frac{ka_k}{sa_0^{1+k/s}}
P_{k-1}^{k/s}(x,s)\right]^{-s/k}.
\end{equation}
The stabilizer $s_k(x)$ is determined from the minimization of the
multiplier defined by
\begin{equation}
\label{13}\mu _k(\varphi,s)=\frac \partial {\partial \varphi}\
y_k(\varphi,s)~,
\end{equation}
which is then transformed into its image
\begin{equation}
\label{14}m_k(x,s)=\mu _k(F_0(x,s),s)
\end{equation}
in terms of the variable $x$, This yields
\begin{equation}
\label{24}m_k(x,s)=\sum_{n=0}^k\frac{a_n}{a_0}\left(1+\frac ns\right)\ x^n ~.
\end{equation}
The control function $s=s_k(x)$ is defined by the equation
\begin{equation}
\label{15}\left| m_k(x,s_k(x))\right| =\min _s\left| m_k(x,s)\right|~ .
\end{equation}
Because the minimization of the multiplier $\left|m_k(x,s)\right|$
makes the trajectory in the space of functions more stable, the role of
the control function $s_k(x)$ is now justified as a stabilizing tool.
This provides finally the self similar approximation (\ref{10}),
\begin{equation}
\label{25}f_k^{*}(x)=p_{k-1}(x)\left[1-\frac{ka_k}{sa_0^{1+k/s}}
x^kp_{k-1}^{k/s}(x)\right]^{-s/k}\ ,
\end{equation}
where $s=s_k(x)$ is the solution of (\ref{15}).

An interesting case occurs when the limit $s \rightarrow \infty$ is taken
for which the approximants will be shown below to be super-exponentials,
i.e. an exponential of an exponential of an exponential...
This limit $s \rightarrow \infty$
often realizes the minimum of the multipliers (hence ensures the
optimal convergence and stability of the procedure) and also corresponds
to the maximum effective order of the
perturbation expansion (\ref{18}).

Note that this limit $s \rightarrow \infty$ does not
always optimize the renormalization procedure at {\it every step}, i.e. does
not always minimize the local multipliers; however, we find that the
solutions are
closed to this limit. In addition, we emphasize that
the fastest convergence of the procedure at each step does not
guarantee that the final expression will converge to the true fixed point.
Quite
often, one can find already at the very first steps of the renormalization that
some multipliers vanish, which mathematically would mean an infinite rate of
convergence, while in practice this makes the procedure trapped
close to a wrong fixed point. It is better to look at the optimal stability
criterion
from a global perspective, i.e. over several steps of the renormalization.
The particularly interesting properties shared by the solutions of the
limit $s \rightarrow \infty$ are the following:
\begin{enumerate}
\item the super-exponentials are obtained under the condition where as many
steps
as possible are performed towards
the sought fixed point. Furthermore, the
total exponentialization is the strongest possible way to renormalize initial
power series.

\item The explicit self-similarity of the final super-exponential
expressions and their
iterative nature allows us to look at them in turn as a sequence of
approximations to the
fixed point and compare their quality with respect to the lowest-order
exponential by means of the ``global'' multipliers. In general,
super-exponentials
surround the fixed point, but do not hit it precisely.

\item For massive numerical calculations it is convenient to have fixed and
reasonable functional forms for the
approximants rather than define them at each step, as
step-by-step renormalizatiion would require.

\end{enumerate}

Taking this limit $s \rightarrow \infty$ in (\ref{25}) gives
\begin{equation}
\label{30}f_k^{*}(x)=p_{k-1}(x)\exp \left(\frac{a_k}{a_0}x^k\right)~.
\end{equation}
Repeating the renormalization, we get
\begin{equation}
\label{31}f_k^{**}(x)=p_{k-2}(x)\exp \{\frac
1{a_0}(a_{k-1}x^{k-1}+a_kx^k)\}.
\end{equation}
Iterating the procedure, we obtain the $k$-fold approximation (\ref{15}) in
the form
\begin{equation}
\label{32}f_k^{*...*}(x)=a_0\exp \{\frac 1{a_0}(a_1x+a_2x^2+...+a_kx^k)\}~.
\end{equation}
We see that the $k$-th approximation (\ref{32}) is
expressed through a part of the initial perturbation series (\ref{17}), namely,
through
$$
p_k(x)-a_0=\sum_{n=1}^ka_nx^n~.
$$
With the notation
\begin{equation}
\label{33}p_k^{^{\prime }}(x)\equiv \sum_{n=0}^ka_n^{^{\prime }}x^n~,
\end{equation}
in which $a_n^{^{\prime }}\equiv a_{n+1,}$ $n=0,1;2,...k,$ we may rewrite
(\ref{32}) as
\begin{equation}
\label{34}f_k^{*...*}(x)=a_0\exp \{\frac x{a_0}p_{k-1}^{^{\prime }}(x)\}~.
\end{equation}

The same functional renormalization procedure can now be applied to the
power series $p_{k-1}^{^{\prime }}(x)$,
giving the corresponding self-similar approximation
\begin{equation}
\label{35}f_{k-1}^{^{\prime }}(x)=a_0^{^{\prime }}\exp \{\frac
x{a_0^{^{\prime }}}p_{k-2}^{^{\prime \prime }}(x)\}~,
\end{equation}
in which
\begin{equation}
\label{36} p_k^{^{\prime \prime }}(x)\equiv \sum_{n=0}^ka_n^{^{\prime
\prime}}x^n,
\quad a_n^{^{\prime \prime }}\equiv a_{n+2} ~ .
\end{equation}
With this renormalization, we transform (\ref{34}) into
\begin{equation}
\label{37}f_k^{*...*}(x)=a_0\exp \{\frac x{a_0}f_{k-1}^{^{\prime }}(x)\}~.
\end{equation}
Combining (\ref{35}) and (\ref{37}), we have
\begin{equation}
\label{38}f_k^{*...*}(x)=a_0\exp \{\frac x{a_0}a_1\exp \{\frac
x{a_1}p_{k-2}^{^{\prime \prime }}(x)\}\}~.
\end{equation}
Converting $k$ times all power series in the exponentials, with the use of
the notation
\begin{equation}
\label{39}b_0=a_{0\ },\quad b_k=\frac{a_k}{a_{k-1}}\ ,\ \ k=1,2,...,
\end{equation}
we obtain the {\it bootstrap self-similar approximation}
\begin{equation}
\label{40}\widetilde{f_k}(x)=b_0\exp (b_1x\exp (b_2x\exp (...b_{k-1}x\exp
(b_kx)))...)~,
\end{equation}
introduced by Yukalov and Gluzman \cite{YG2}. It is this fundamental result
that we use to obtain expressions (\ref{sen1}) and (\ref{sen2}) in the main
text.

\pagebreak

\begin{figure}
\begin{center}
\caption{\protect\label{fig1} US dollar in German mark and Japanese Yen
from Jan. 4, 1971 till May 19, 1999.
The plots at the top show the time series of the prices. The middle plots show
the Froude number defined
in equation (\ref{Froude}) as a function of the reduced prediction horizon
$\Delta T \equiv (A_1/A_0) \delta t$ where $\delta t$ is fixed equal to 5
days.
The plots at the bottom
show the number of realizations of each of the six relevant patterns as a
function of
the threshold for the predicted amplitude of the price move. The symbols are
$p_1 \equiv \rfloor +$  (x), $p_3 \equiv \lceil +$ (+), $p_4 \equiv \lceil
-$ (o),
$p_6 \equiv \rceil -$ (.), $p_7 \equiv \lfloor +$ (square) and $p_8 \equiv
\lfloor -$
(diamond). The dotted, dashed and continous lines delineate domains of
different predicted
return signs (see text).
}
\end{center}
\end{figure}

%\pagebreak

\begin{figure}
\begin{center}
\caption{\protect\label{fig2} Same as figure \ref{fig1} for the British
pound and the Swiss franc.
}
\end{center}
\end{figure}

%\pagebreak

\begin{figure}
\begin{center}
\caption{\protect\label{fig3} Same as figure \ref{fig1} for the thirty year
US treasury bond
TYX from Oct. 29, 1993 till  Aug. 9, 1999, and the Japanese Governement
Bond JGB
 from from Jan. 1, 1992 till March 23, 1999.
}
\end{center}
\end{figure}

%\pagebreak

\begin{figure}
\begin{center}
\caption{\protect\label{fig4} Same as figure \ref{fig1} for the
 SP500 index from Jan. 1, 1950 till June 1, 1999, and
 the Dow Jones index from Jan. 2, 1970 till Feb. 24, 1998.
}
\end{center}
\end{figure}

%\pagebreak

\begin{figure}
\begin{center}
\caption{\protect\label{fig5} Same as figure \ref{fig1} for the
 Nasdaq index from Feb. 5, 1971 till May 18, 1999, and the
 Japanese Nikkei index from April 16, 1990 till May 17, 1999.
}
\end{center}
\end{figure}

%\pagebreak

\begin{figure}
\begin{center}
\caption{\protect\label{fig6} Same as figure \ref{fig1} for the
British FTSE index from April 17, 1990 till May 17, 1999, and
the German DAX index from July 1, 1991 till May 17, 1999.
}
\end{center}
\end{figure}

%\pagebreak

\begin{figure}
\begin{center}
\caption{\protect\label{figsupbull1} }
% \pagebreak
%\vspace{1cm}
%\epsfig{file=figsupbull-b.eps,height=16cm,width=16cm}
\caption{\protect\label{figsupbull2} Super bull case $\rfloor$ : comparative
statistical tests of the predictions of our theory
(average approximant $S^*$) represented by crosses, a trend following
strategy (linear approximation of $S_1$)
represented by open squares and
the bare parabolic parameterization (\ref{nhakak}) represented by open circles,
 for the twelve assets presented in figure \ref{fig1}-\ref{fig6}.
 For information on how the figures are constructed, see the main text.}
\end{center}
\end{figure}

%\pagebreak

\begin{figure}
\begin{center}
\caption{\protect\label{FigBalbull} ``Phase diagram'' for the balanced-bull
regime $\lceil$,
defined in the
parameter space
$(\Delta T/T_N, |F|)$ with $T_N = 1$, delineating the regions of
positive and negative returns. The boundaries $F_1(\Delta T)$,
$F_2(\Delta T)$ and $F_3(\Delta T)$ are shown as dotted-dashed, dashed and
continuous lines
respectively. The long-dashed line indicated as ``regression'' on the figure
correspondsto the solution of ${S_0(T_N+\Delta T,F)-{S_0(T_N,F)=0,}}$.}
\end{center}
\end{figure}

%\pagebreak

\begin{figure}
\begin{center}
\caption{\protect\label{FigBalbullret} Predicted return $R$ in the
balanced-bull
regime $\lceil$ defined by equation
(\ref{fnalnacl}) as a function of the ratio $\Delta T/ T_N$ of the
prediction horizon
over the learning interval for several values of the Froude number.
}
\end{center}
\end{figure}

%\pagebreak

\begin{figure}
\begin{center}
\caption{\protect\label{figp3p4testa} }
% \pagebreak
% \epsfig{file=figp3p4test-b.eps,height=16cm,width=16cm}
\caption{\protect\label{figp3p4testb} }
% \pagebreak
% \epsfig{file=figp3p4test-c.eps,height=16cm,width=16cm}
\caption{\protect\label{figp3p4testc} }
% \pagebreak
%\epsfig{file=figp3p4test-d.eps,height=16cm,width=16cm}
\caption{\protect\label{figp3p4testd}  Balanced bull case $\lceil$:
Comparative statistical tests of the predictions of our theory
(average approximant $S^*$) represented by crosses, a trend following
strategy (linear approximation of $S_1$)
represented by open squares and
the bare parabolic parameterization (\ref{nhakak}) represented by open circles,
 for the twelve assets presented in figure \ref{fig1}-\ref{fig6}. The
predictions of
 positive $\lceil +$ and negative $\lceil -$ returns are represented
separately.
The same parameters as for the super bull regime have been used
(learning interval is fixed to 15 days and the prediction horizon is
fixed to 5 days). The same quantities as in figure
\ref{figsupbull1},\ref{figsupbull2} are
represented.
 For information on how the figures are constructed, see the main text.
}
\end{center}
\end{figure}

%\pagebreak

\begin{figure}
\begin{center}
\caption{\protect\label{figsupbear1} }
% \pagebreak
%\vspace{1cm}
%\epsfig{file=figsupbear-b.eps,height=16cm,width=16cm}
\caption{\protect\label{figsupbear2}
Super-bear $\rceil$: comparative
statistical tests
of the predictions of our theory
(average approximant $S^*$) represented by crosses, a trend following
strategy (linear approximation of $S_1$)
represented by open squares and
the bare parabolic parameterization (\ref{nhakak}) represented by open circles,
 for the twelve assets presented in figure \ref{fig1}-\ref{fig6}.
 For information on how the figures are constructed, see the main text.
}
\end{center}
\end{figure}

%\pagebreak

\begin{figure}
\begin{center}
\caption{\protect\label{FigBalbear} ``Phase diagram'' of the balanced-bear
regime $\lfloor$, defined in the
parameter space
$(\Delta T/T_N, |F|)$ with $T_N = 1$, delineating the regions of
positive and negative returns. The boundaries $F_1(\Delta T)$,
$F_2(\Delta T)$ and $F_3(\Delta T)$ are shown as dashed, continuous and
dotted lines
respectively. The dotted-dashed line indicated as ``regression'' on the
figure corresponds
to the solution of ${S_0(T_N+\Delta T,F)-{S_0(T_N,F)=0,}}$.
}
\end{center}
\end{figure}

%\pagebreak

\begin{figure}
\begin{center}
\caption{\protect\label{FigBalbearret} Predicted return $R$
of the balanced-bear
regime $\lfloor$, defined by equation
(\ref{fnalnacl}), as a function of the ratio $\Delta T/ T_N$ of the
prediction horizon
over the learning interval for several values of the Froude number.
}
\end{center}
\end{figure}

%\pagebreak

\begin{figure}
\begin{center}
\caption{\protect\label{figp7p8testa} }
% \pagebreak
% \epsfig{file=figp7p8test-b.eps,height=16cm,width=16cm}
\caption{\protect\label{figp7p8testb} }
% \pagebreak
% \epsfig{file=figp7p8test-c.eps,height=16cm,width=16cm}
\caption{\protect\label{figp7p8testc} }
% \pagebreak
%\epsfig{file=figp7p8test-d.eps,height=16cm,width=16cm}
\caption{\protect\label{figp7p8testd}  Balanced bear case $\lfloor$:
Comparative statistical tests of the predictions of our theory
(average approximant $S^*$) represented by crosses, a trend following
strategy (linear approximation of $S_1$)
represented by open squares and
the bare parabolic parameterization (\ref{nhakak}) represented by open circles,
 for the twelve assets presented in figure \ref{fig1}-\ref{fig6}. The
predictions of
 positive $\lfloor +$ (p7) and negative $\lfloor -$ (p8) returns are
represented separately.
The same parameters as for the super bull regime have been used
(learning interval is fixed to 15 days and the prediction horizon is
fixed to 5 days). The same quantities as in figure
\ref{figsupbull1},\ref{figsupbull2} are
represented.
 For information on how the figures are constructed, see the main text.
}
\end{center}
\end{figure}

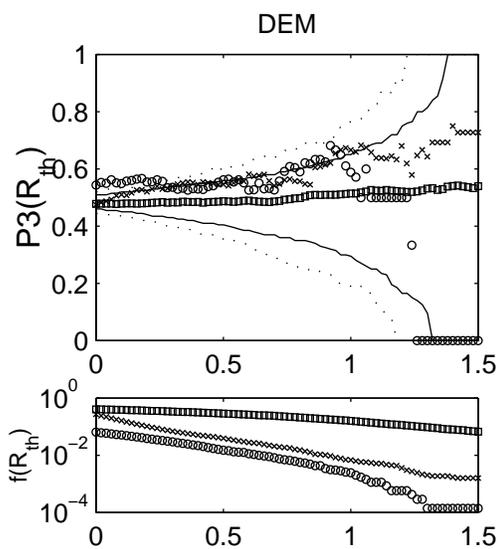
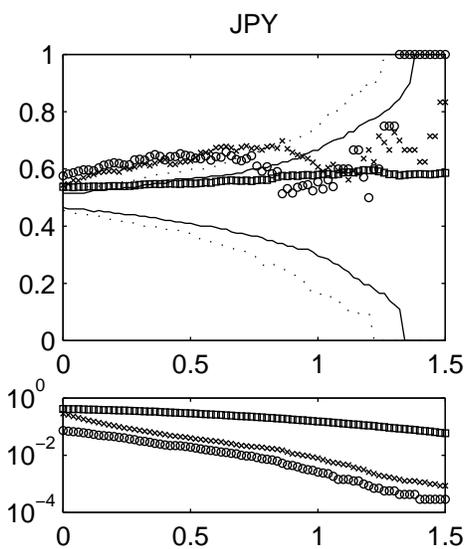
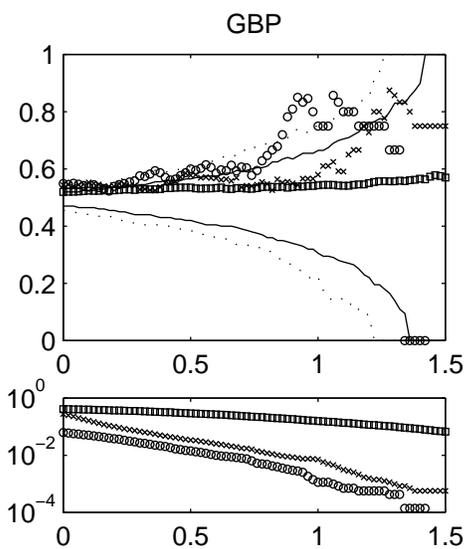
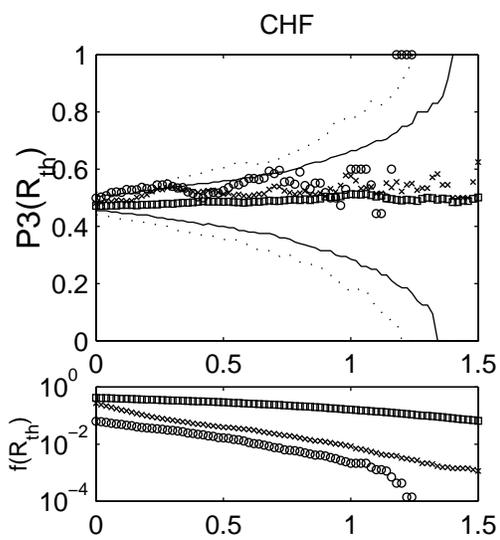
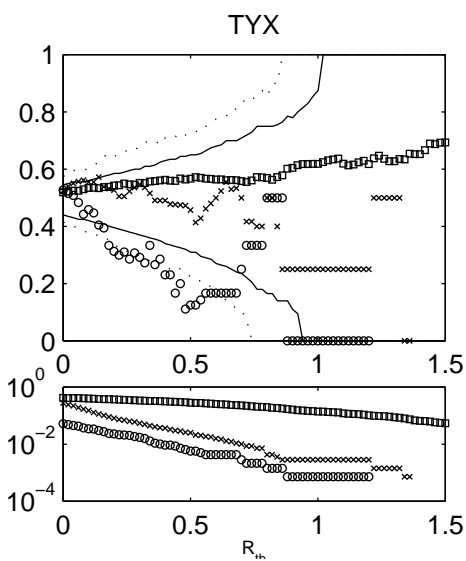
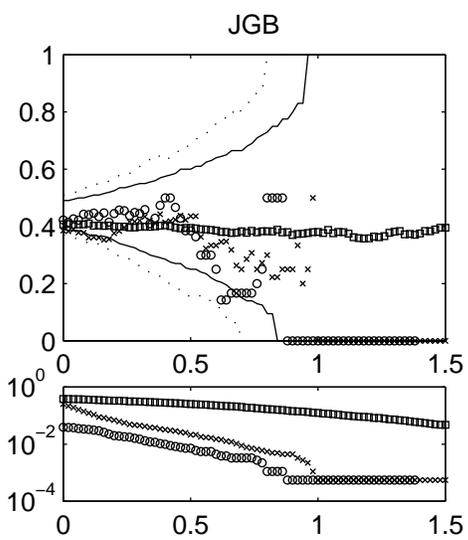
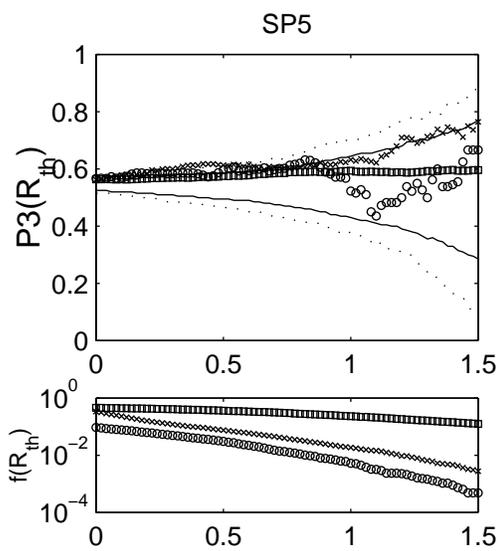
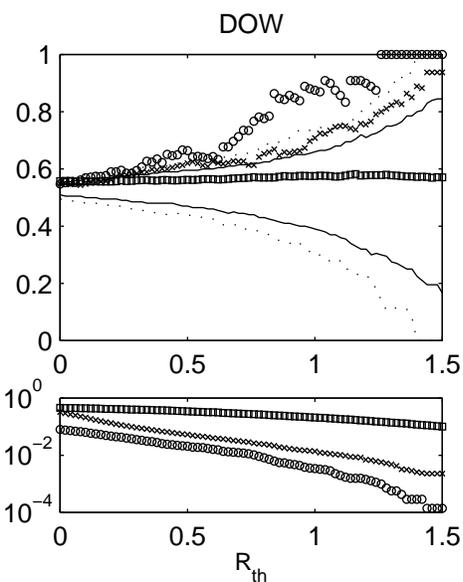
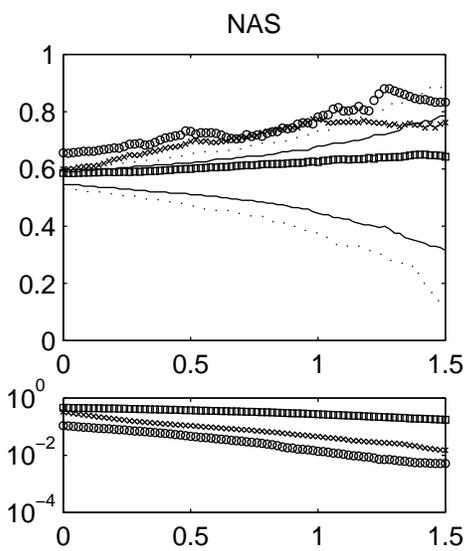
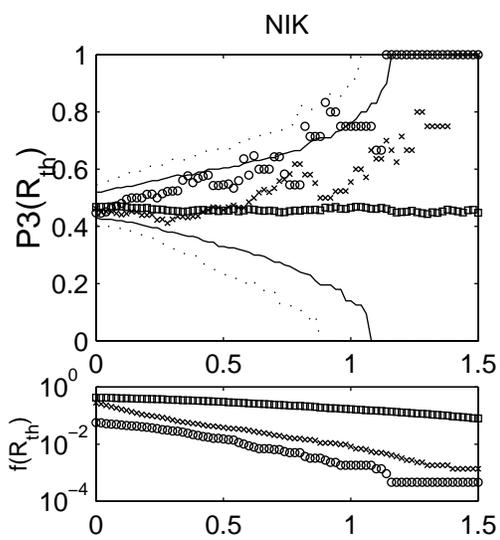
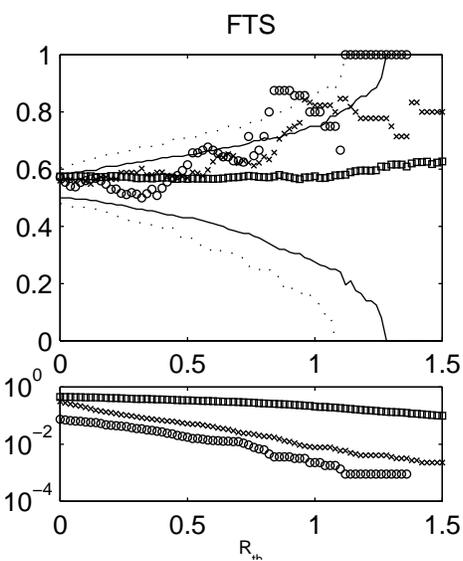
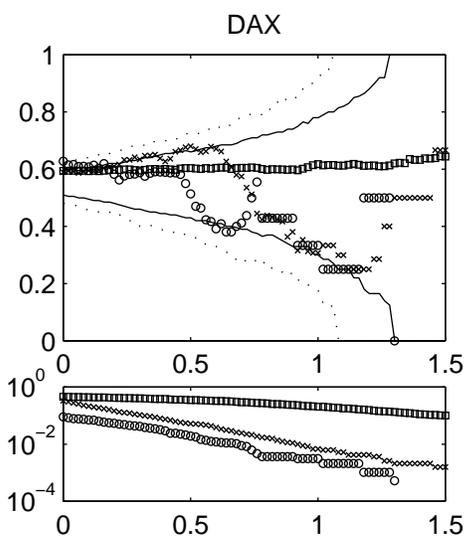
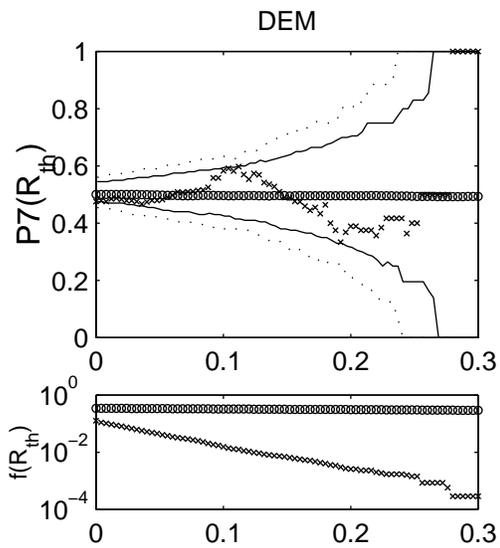
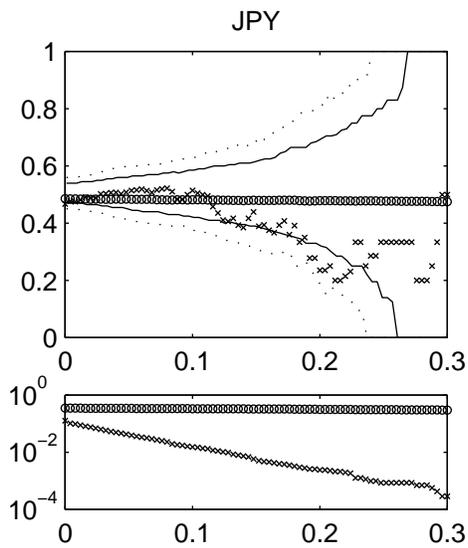
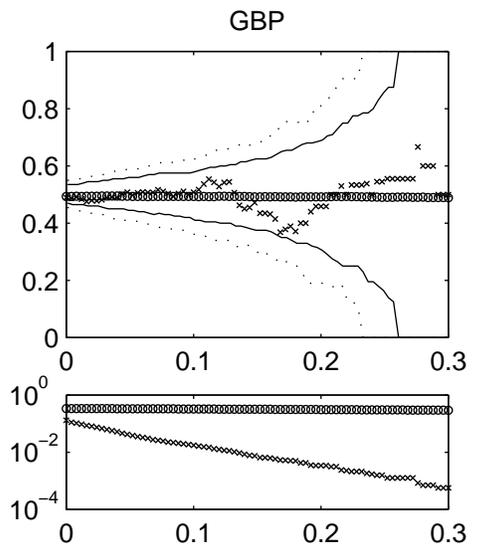
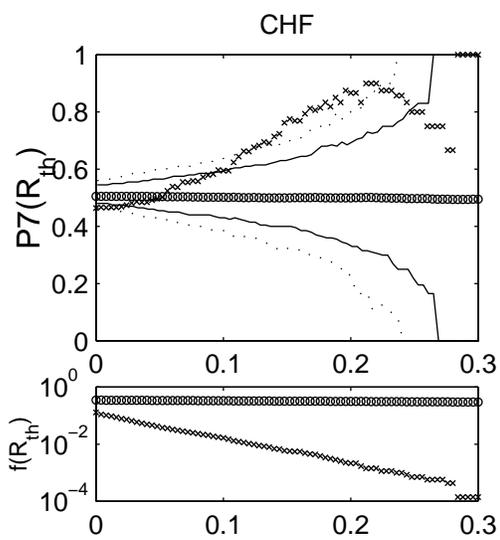
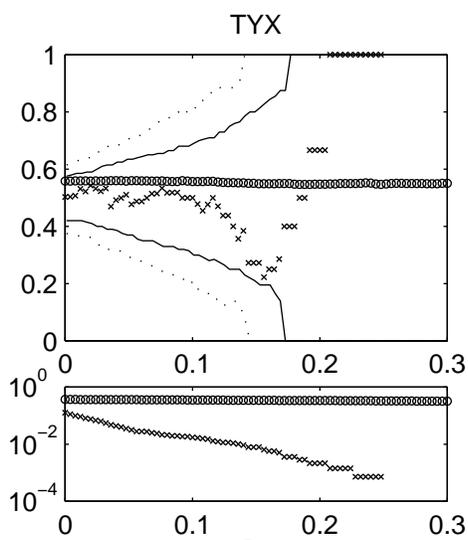
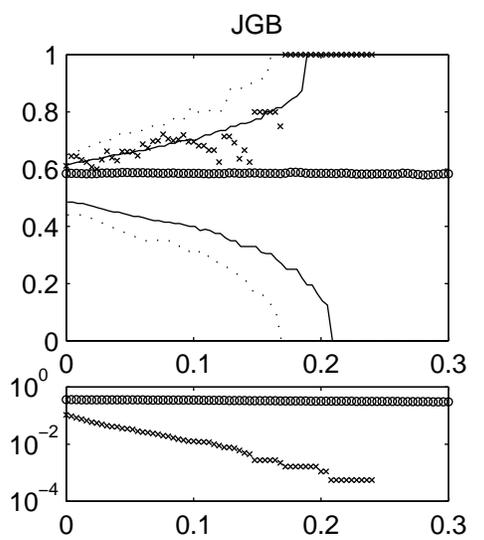
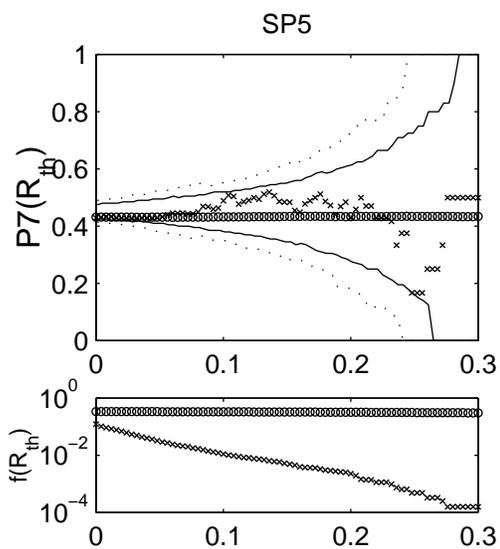
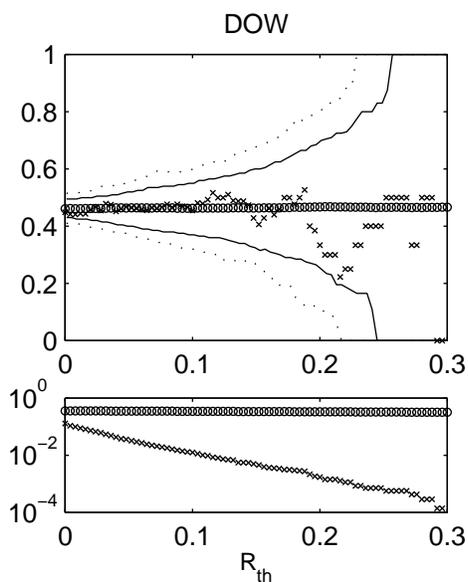
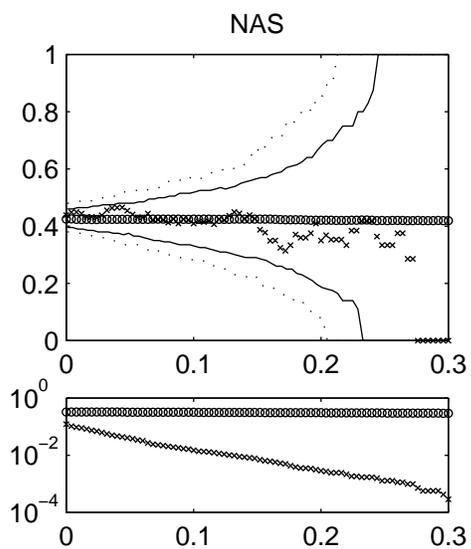
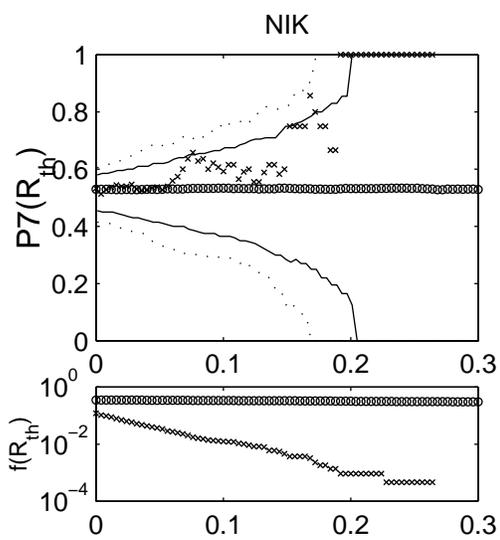
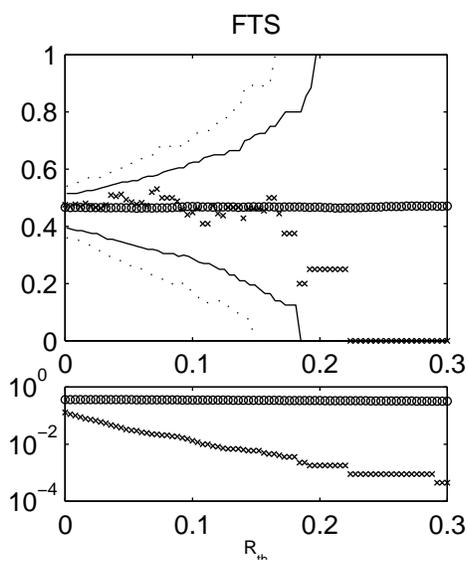
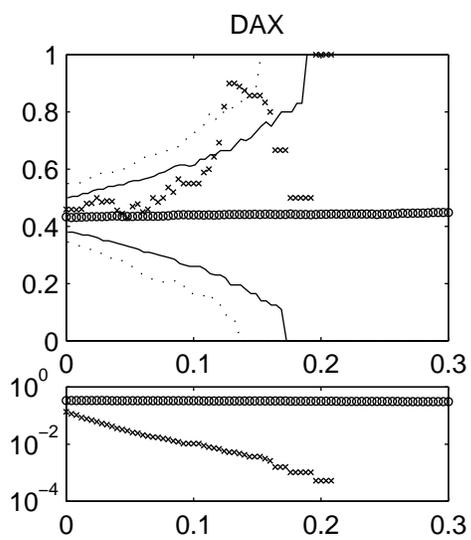
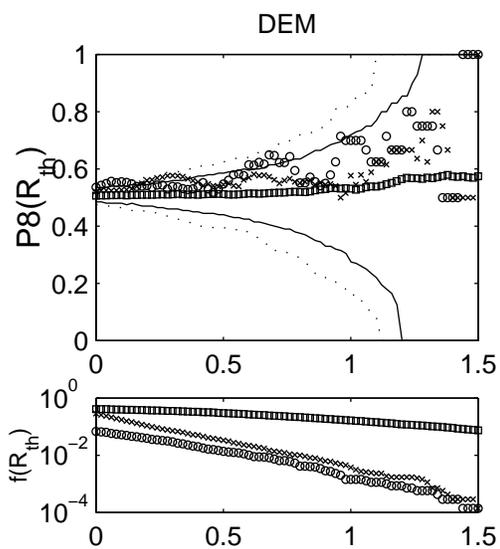
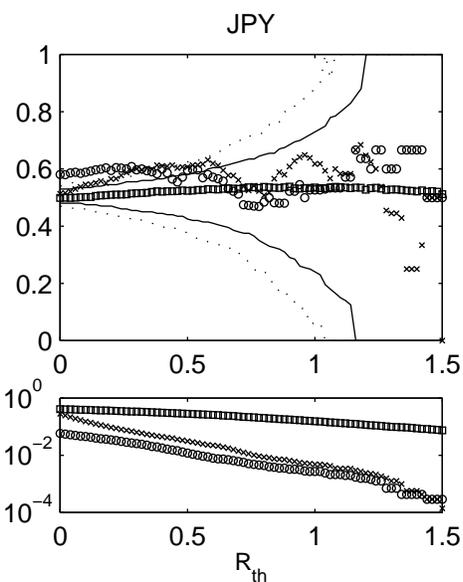
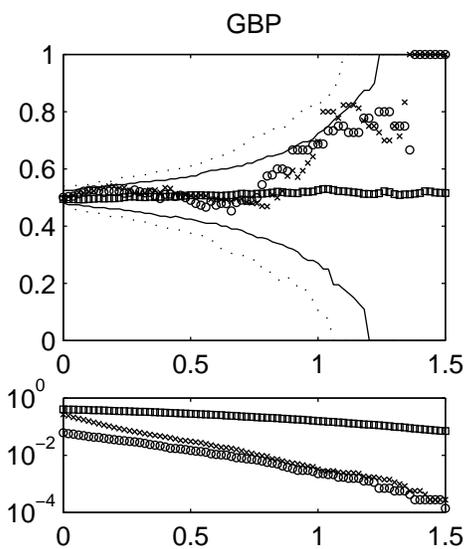
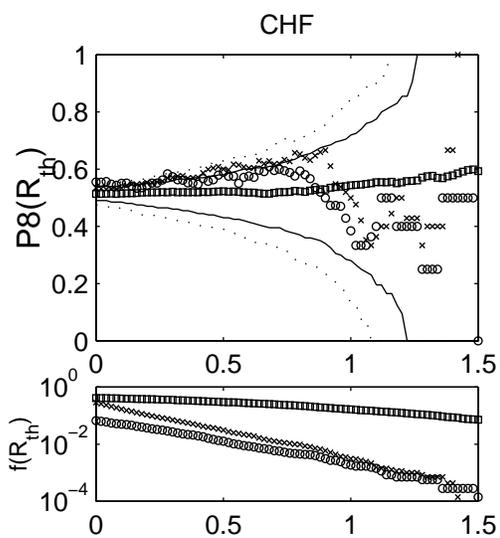
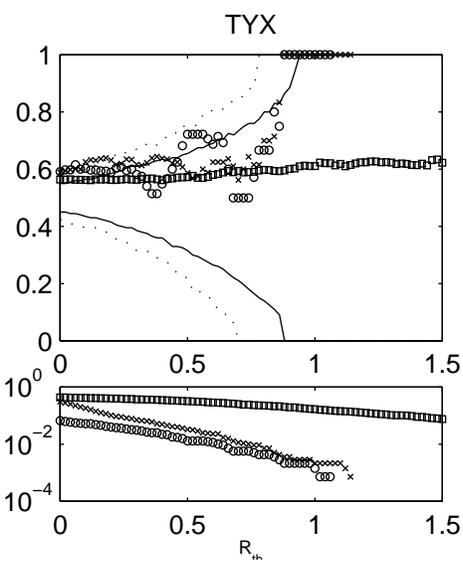
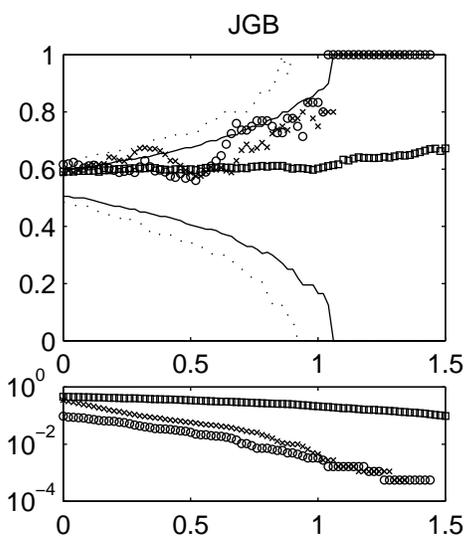
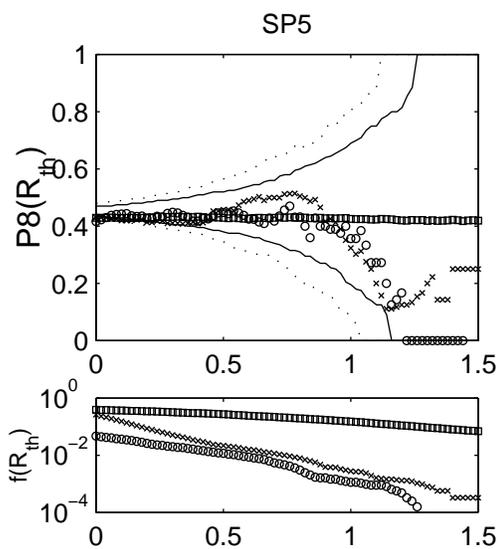
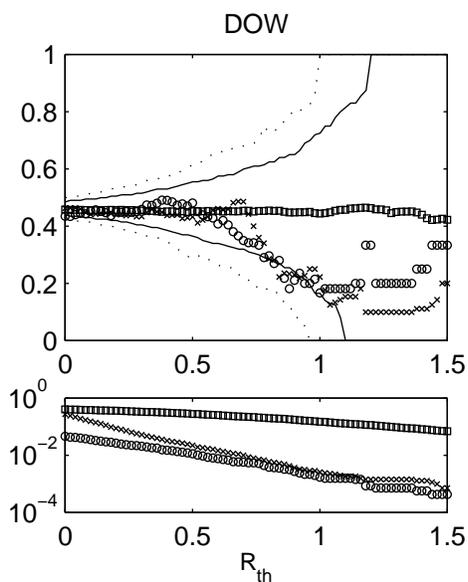
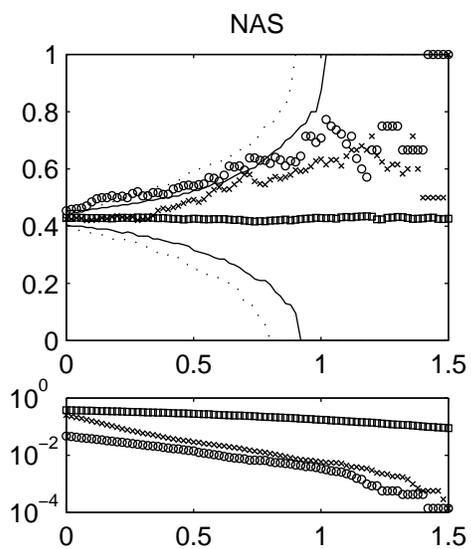
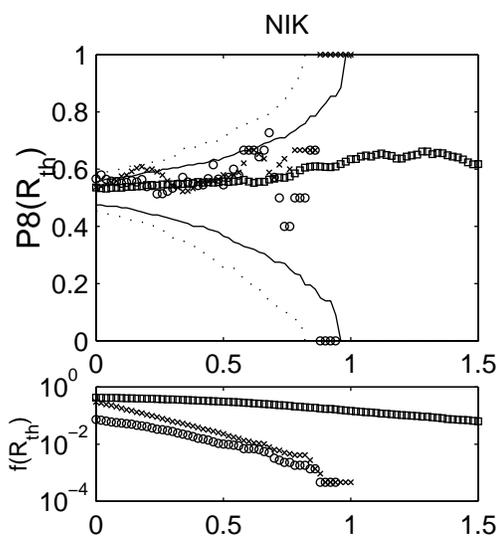
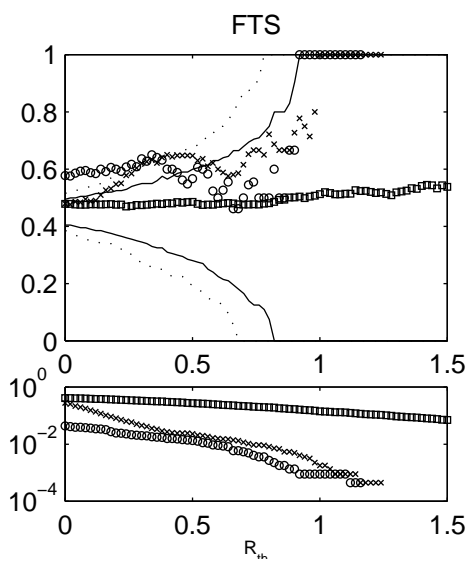
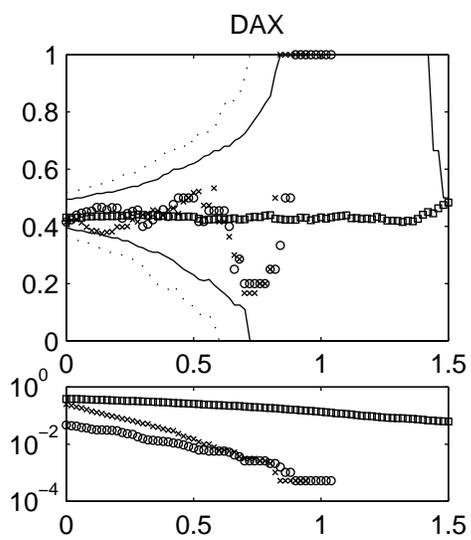
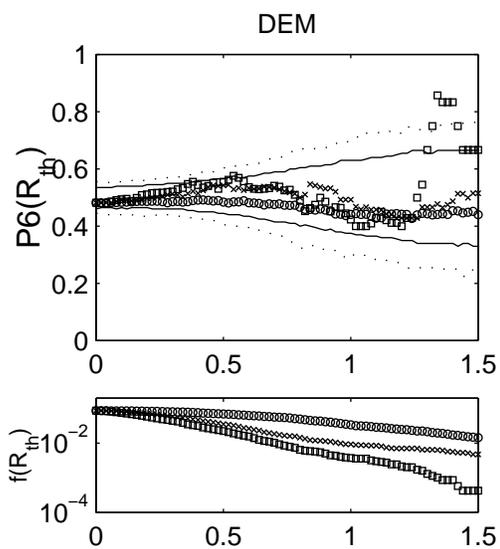
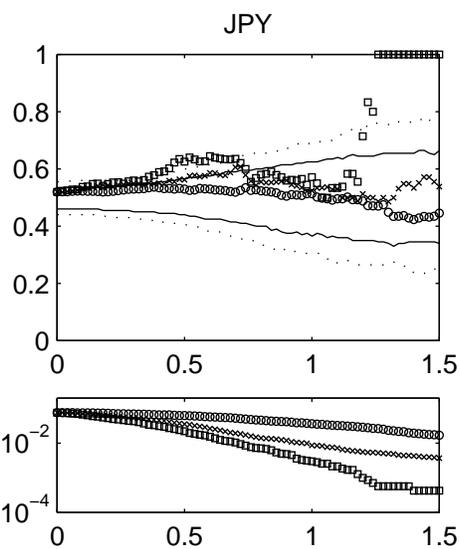
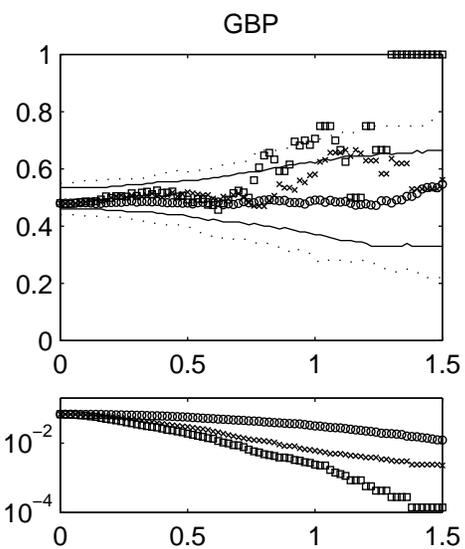
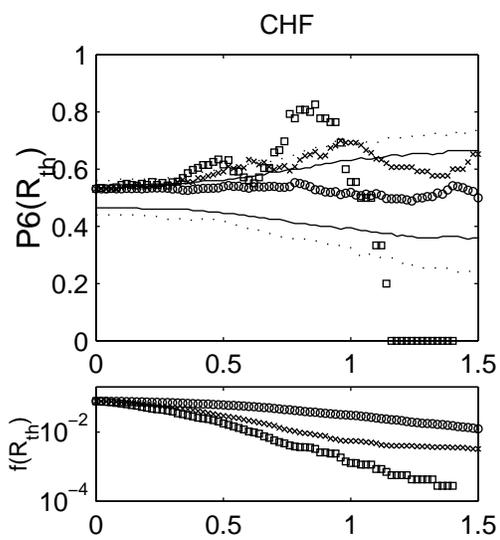
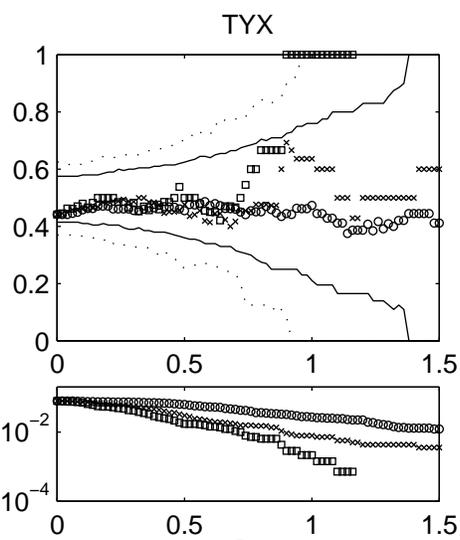
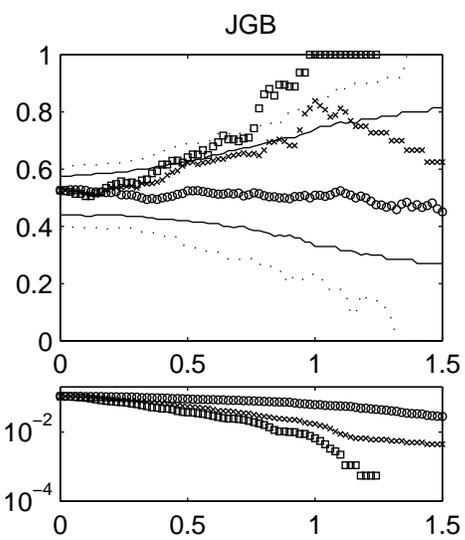
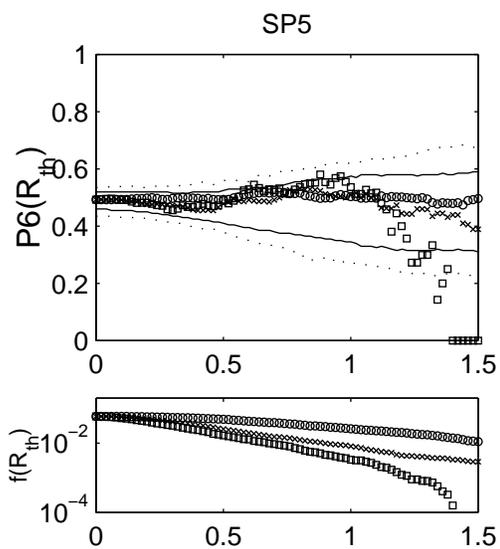
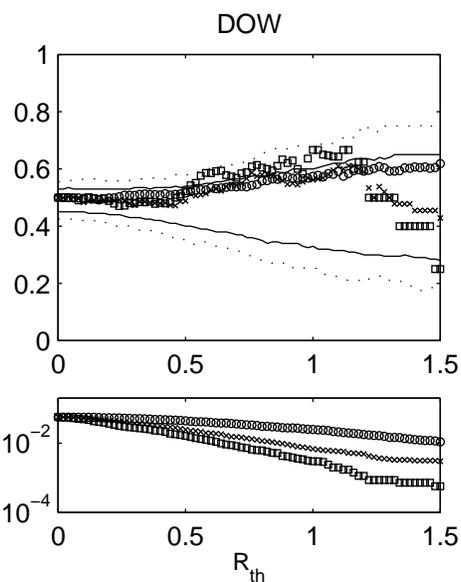
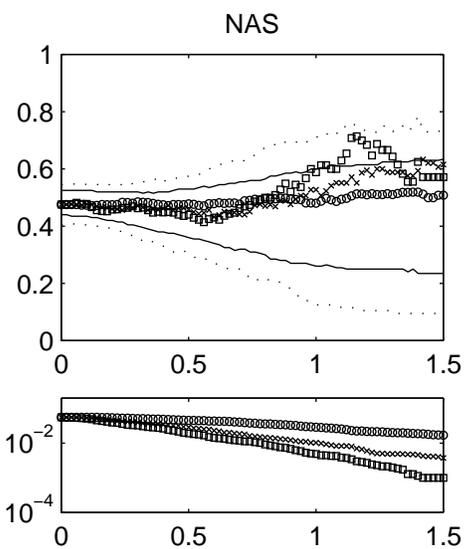
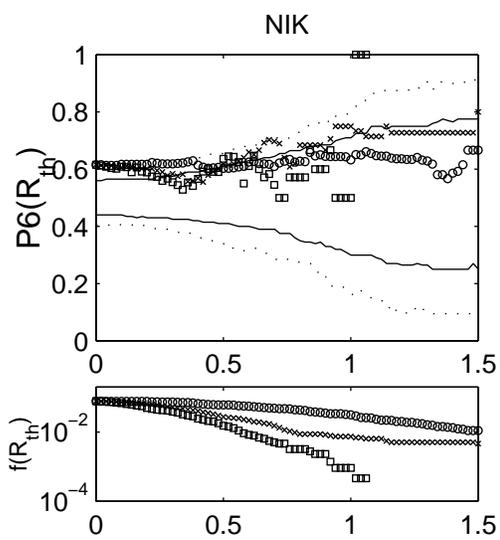
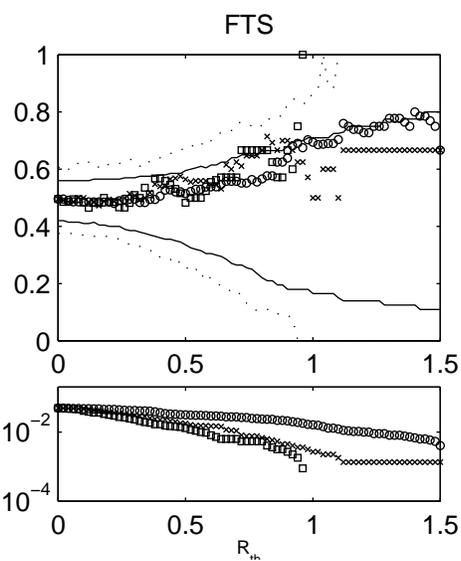
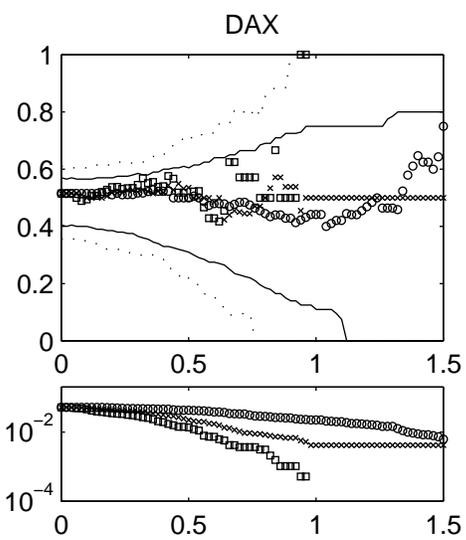
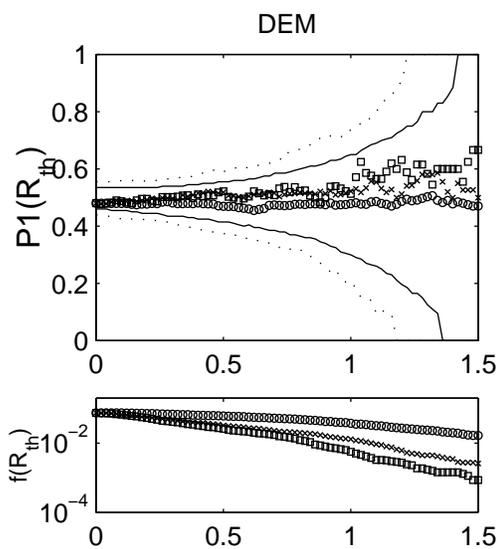
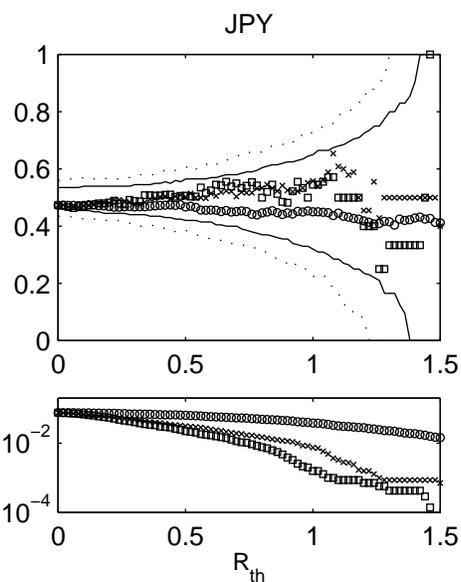
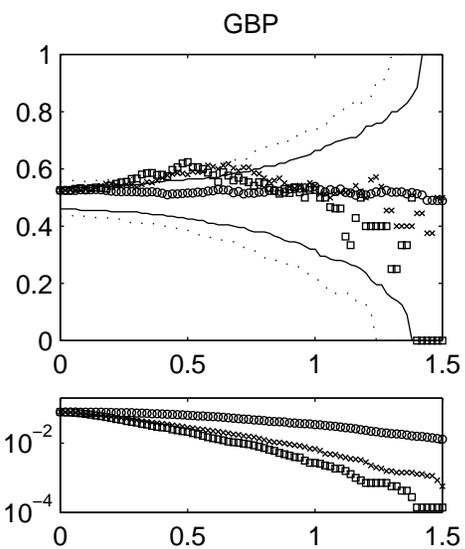
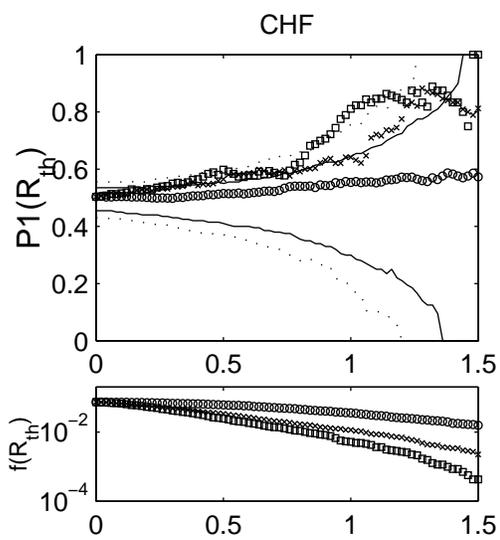
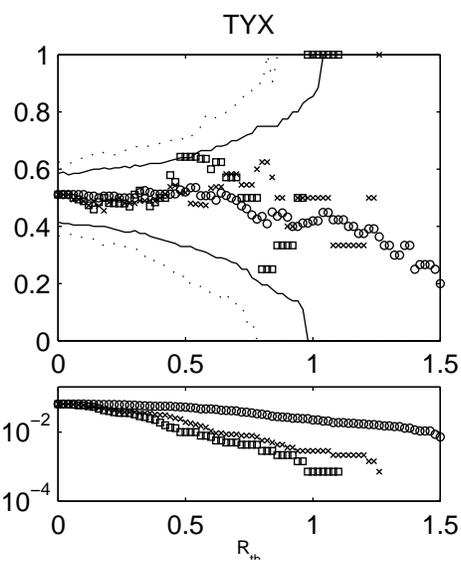
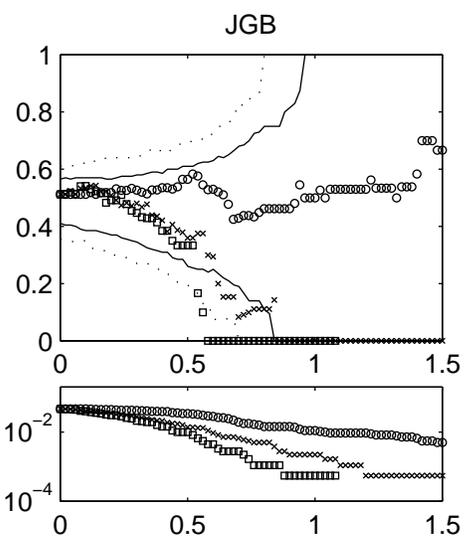
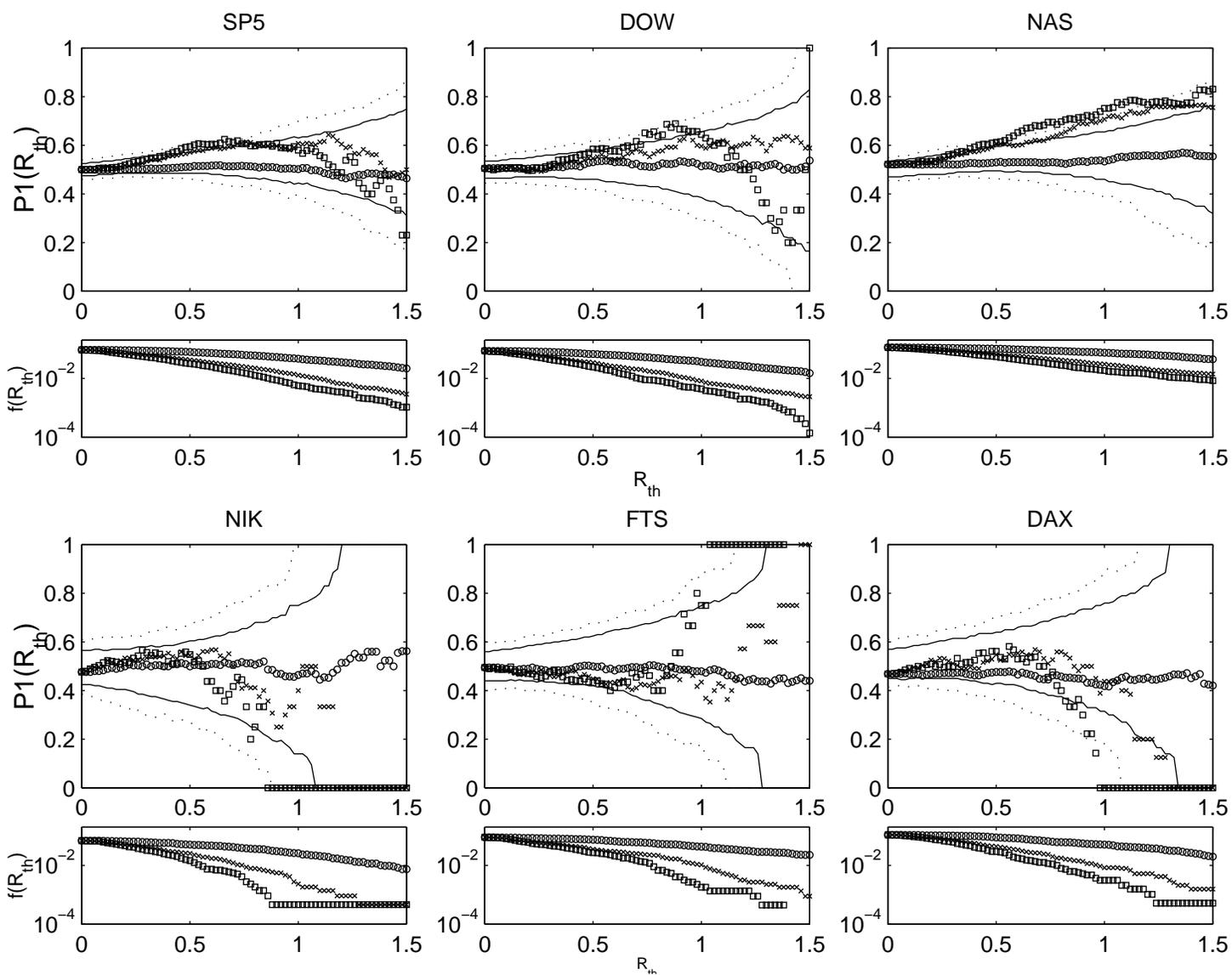
\end{document}